\documentclass[prb,twocolumn,amssymb,showpacs]{revtex4}
\usepackage{graphicx,amsmath,mathrsfs}
\begin{document}

\title{Charge-spin duality in non-equilibrium transport of helical liquids}

\author{Chao-Xing Liu$^{1,2}$, Jan Carl Budich$^{1}$,  Patrik Recher$^1$, and Bj${\rm \ddot{o}}$rn Trauzettel$^1$ }

\affiliation{ $^1$Institute for Theoretical Physics and Astrophysics,
University of W$\ddot{u}$rzburg, 97074 W$\ddot{u}$rzburg, Germany;\\
 $^2$Physikalisches Institut (EP3), University of W$\ddot{u}$rzburg, 97074 W$\ddot{u}$rzburg, Germany; }

\date\today

\begin{abstract}
Non-equilibrium transport properties of charge and spin sector of two edges of a quantum spin Hall insulator are investigated theoretically in a four-terminal configuration. A simple duality relation between charge and spin sector
is found for two helical Tomonaga Luttinger liquids (hTTLs) connected to non-interacting electron reservoirs. If the hTLLs on opposite edges are coupled locally or non-locally, the mixing between them yields interesting physics where spin information can be easily detected by a charge measurement and vice versa. Particularly, we show how a pure spin density in the absence of charge current can be generated in a setup that contains two hTLL and one spinful Tomonaga Luttinger liquid in between.
\end{abstract}


\maketitle

{\it Introduction.--}
The discovery of topological insulators (TIs) in both two spatial dimensions (2D)
and three spatial dimensions (3D) has recently attracted a lot of interest \cite{qi2010,moore2010,hasan2010}.
Unlike in normal insulators, there is a gapless mode appearing within the bulk gap
at the edge of TIs which originates from strong spin-orbit coupling and
is protected by time reversal. In 2D, a TI is also called
quantum spin Hall (QSH) insulator since its edge states are one dimensional (1D)
counter-propagating modes with opposite spin. These 1D systems have been coined helical liquids or helical Tomonaga Luttinger liquids (hTLLs).
Transport properties of hTLLs have been predicted and observed
at the edge of HgTe quantum wells \cite{bernevig2006d,koenig2007}
and proposed to also exist in InAs/GaSb quantum wells \cite{liu2008} as well as
Bi$_2$Se$_3$ or Bi$_2$Te$_3$ thin films \cite{lu2010,liu2010}. An important feature of the hTLL is that spin and momentum are locked to each other.
Remarkably, one hTLL has only half the degrees of freedom of a spinful Tomonaga Luttinger liquid (sTLL). Thus, two hTLLs,
which naturally exist at two opposite edges of a QSH insulator, can recover the degrees of freedom of a sTLL. It is well known and has even been experimentally confirmed that there is spin-charge separation for a 1D sTLL \cite{auslaender2005}.

Therefore it is natural to ask the question how spin and charge sector behave for two uncoupled as well as two coupled hTLLs. In this Letter, we investigate the non-equilibrium transport properties of two hTLLs in a four-terminal configuration. Most interestingly, we find a duality relation between charge and spin sector of two hTLLs taking into
account the coupling to non-interacting electron reservoirs. As a physical consequence, there is a simple relation
between charge current and spin polarization in the dual voltage configurations of two hTLLs (see below).
Importantly, the coupling between two edges will destroy the simple duality relation. However, we can still
manipulate the charge and spin sector separately only by electric means. To demonstrate this, we study different scattering mechanisms between the two hTLLs within the non-equilibrium Keldysh formalism and bosonization. Different bias dependencies are found for different scattering mechanisms which can be used to distinguish and identify them in experiments.

\begin{figure}[h]
\begin{center}
\includegraphics[width=3.2in]{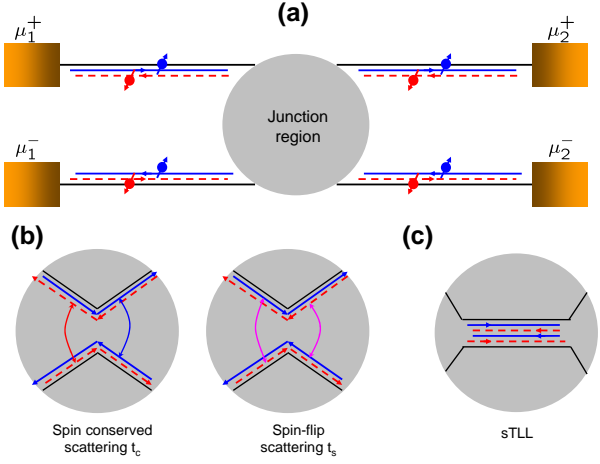}
\end{center}
\caption{(Color online) (a) Schematic of the four-terminal setup. At each edge, there is
a conducting channel of a hTLL (blue full lines correspond to spin up; red dashed lines to spin down). The two hTLLs are mixed in the junction region and different types of junctions are analyzed:
(b) the short junction with two possible single particle scattering terms:
(i) spin conserved scattering $t_c$ and (ii) spin-flip scattering $t_s$ and
(c) the long junction modeled by a sTLL. }
\label{fig:configuration}
\end{figure}

{\it Model and spin-charge duality.--} We consider a QSH insulator in a four-terminal
configuration as shown in Fig.~\ref{fig:configuration} (a).
The two edges are denoted by $\alpha=+(-)$ for the upper (lower)
edge. On each edge $\alpha$, there are two terminals with chemical potentials
$\mu_{i,\alpha}$ ($i=1,2$ means left and right lead, respectively). Two terminals on the same edge are connected by a hTLL of finite length $L$. The hTLL states are described by
field operators $\psi_{a\sigma}$ where $(a,\sigma)=(R,\uparrow)$
or $(L,\downarrow)$ for the upper edge and $(a,\sigma)=(R,\downarrow)$
or $(L,\uparrow)$ for the lower edge. In the middle region of the sample, the hTLLs at the two edges can mix and different types of coupling mechanisms will be discussed below.

Interestingly, there are two possibilities for choosing the basis states of the system: the helical edge basis and the spin-charge basis. For the helical edge basis,
the non-chiral boson field is defined separately for each edge with
$\varphi_{+(-)}=\phi_{R\uparrow(\downarrow)}+\phi_{L\downarrow(\uparrow)}$
at the upper (lower) edge and the corresponding dual field
$\theta_{+(-)}=\phi_{R\uparrow(\downarrow)}-\phi_{L\downarrow(\uparrow)}$. Here,
$\phi_{a\sigma}$ ($a=R,L$ and $\sigma=\uparrow,\downarrow$) is
the standard boson field operator in bosonization \cite{giamarchi2004}.
This basis is suitable to study the current at different terminals.
However, when we are interested in spin properties, it is more convenient to introduce
the spin-charge basis, which is related to the helical edge basis by
\begin{eqnarray}
    &&\varphi_c=\frac{1}{\sqrt{2}}(\varphi_++\varphi_-),\qquad
    \varphi_s=\frac{1}{\sqrt{2}}(\theta_+-\theta_-),\nonumber\\
    &&\theta_c=\frac{1}{\sqrt{2}}(\theta_++\theta_-),\qquad
    \theta_s=\frac{1}{\sqrt{2}}(\varphi_+-\varphi_-),
    \label{eq:tran1}
\end{eqnarray}
where $c$ and $s$ represent charge and spin sector, respectively. The Hamiltonian can be written as
\begin{eqnarray}
    \hat{H}=\hat{H}_0+\hat{H}_V+\hat{H}_t,
    \label{eq:Ham}
\end{eqnarray}
where $\hat{H}_0$ describes the hTLLs at two edges,
$\hat{H}_V$ the coupling between the helical liquid and
the leads, and $\hat{H}_t$ the scattering region.

The hTLL coupled to electron reservoirs can be modeled by the so-called $g(x)$-model \cite{maslov1995,ponomarenko1995,safi1995} given by
\begin{eqnarray}
    H^c_{0}=\frac{\hbar v_f}{2}\int dx\left[ (\partial_x\theta_c)^2+\frac{1}{g^2(x)}
    (\partial_x\varphi_c)^2\right]\nonumber\\
    H^s_{0}=\frac{\hbar v_f}{2}\int dx\left[ \frac{1}{g^2(x)}(\partial_x\theta_s)^2+
    (\partial_x\varphi_s)^2\right],
    \label{eq:H0hl2}
\end{eqnarray}
in the spin-charge basis. Here, $v_f$ is the Fermi velocity and
$g(x)$ is the hTLL interaction parameter ($g(x)=g_0<1$ for repulsive interactions within the helical edge located at $|x|<L/2$, and $g(x)=1$ for the non-interacting fermions in the leads with $|x|>L/2$) \footnote{We assume the same parameters $v_f$ and $g_0$ at the two edges ($\alpha=\pm$). This is reasonable for two edges of the same system.}. The chemical potentials in the leads are naturally
taken into account with the Hamiltonian
$H_V=\int \frac{dx}{\sqrt{2\pi}}\left[ \partial_x\mu_c(x)\varphi_c
    +\partial_x\mu_s(x)\theta_s \right]$
where $\partial_x\mu_{c(s)}=-\mu_{1,c(s)}\delta(x+L/2)+\mu_{2,c(s)}\delta(x-L/2)$
with $\mu_{i,c}=\mu_{i,+}+\mu_{i,-}$ and $\mu_{i,s}=\mu_{i,+}-\mu_{i,-}$
($i=1,2$).
Remarkably, $\mu_c$ couples to $\varphi_c$ while
$\mu_s$ couples to $\theta_s$. Therefore, the electric voltage can couple
to both charge and spin sector.
This provides us an easy way to control charge and spin separately -- in contrast to the usual sTLL where the electric chemical potential only couples to the charge sector.
Moreover, we discover that there is a duality relation between charge and spin sector, namely
\begin{eqnarray}
	\varphi_c\leftrightarrow\theta_s\qquad
        \theta_c\leftrightarrow\varphi_s.
	\label{eq:duality}
\end{eqnarray}
For $\hat{H}^\lambda_0$ ($\lambda=c,s$), the above duality relation
is directly related to the constraint $g_c = 1/g_s$ discovered in Ref. [\onlinecite{hou2009}] before.
Here, we show that this relation remains valid even if the system is coupled to biased electron reservoirs. Thus, the duality relation should be observable in transport properties of the system.

\begin{figure}[h]
\begin{center}
\includegraphics[width=2.8in]{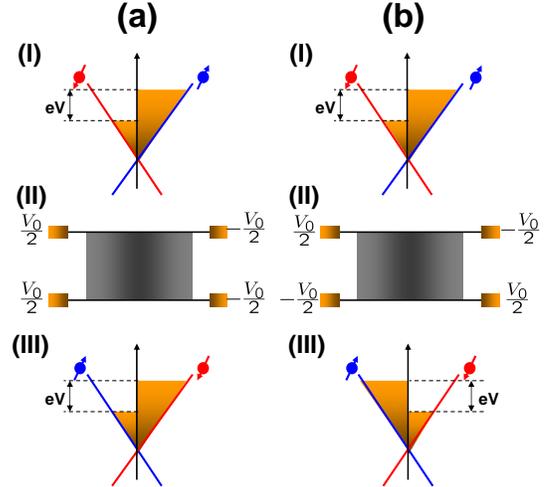}
\end{center}
\caption{ (Color online) (a). (I) and (III) show band dispersions and chemical potentials of the upper edge and the lower edge, respectively.
(II) illustrates the voltage configuration $\mu_{1,+}=\mu_{1,-}=eV_0/2$ and
$\mu_{2,+}=\mu_{2,-}=-eV_0/2$, yielding $V_c=V_0$ and $V_s=0$;
(b). Similar to (a) with a different voltage configuration $\mu_{1,+}=\mu_{2,-}=eV_0/2$
and $\mu_{2,+}=\mu_{1,-}=-eV_0/2$, giving $V_c=0$ and $V_s=V_0$.
Note that in (a) there is a finite charge current but no spin density while
in (b) there is a finite spin density but no charge current.}\label{fig:duality}
\end{figure}

What is the physical consequence of this duality relation? To answer this question, we investigate the total charge current
and spin density of the system.
The charge current is given by $\hat{j}_c=e\sqrt{\frac{2}{\pi}}
\partial_t\varphi_c$, which is the sum of the currents along the two edges
$\hat{j}_c=\hat{j}_++\hat{j}_-$, with $\hat{j}_\alpha(x)=\frac{e}{\sqrt{\pi}}
\partial_t\varphi_\alpha$. The spin density can be defined as
$\hat{\rho}_s=\sqrt{\frac{2}{\pi}}\partial_x\varphi_s$. Combining Eq.~(\ref{eq:tran1}) and $\partial_t\varphi_\alpha=-v_f\partial_x\theta_\alpha$,
it is evident that the spin density can be directly related
to the charge current along the two edges by
$\hat{\rho}_s=-\frac{1}{ev_f}(\hat{j}_+-\hat{j}_-)$.
In the absence of mixing between two edges, our setup describes transport through two independent 1D channels.
Then, it follows directly from previous work \cite{maslov1995,ponomarenko1995,safi1995} that $\langle\hat{j}_{0,\alpha}\rangle
=\frac{e^2}{h}V_\alpha$ where $eV_\alpha=\mu_{1,\alpha}-\mu_{2,\alpha}$.
Thus, the total charge current is $\langle\hat{j}_{0,c}\rangle=\frac{2e^2}{h}V_c$
with $V_c=\frac{V_++V_-}{2}=\frac{1}{2e}(\mu_{1,c}-\mu_{2,c})$, while the total spin density
is given by $\langle\hat{\rho}_{0,s}\rangle=-\frac{2e}{hv_f}V_s$ with
$V_s=\frac{V_+-V_-}{2}=\frac{1}{2e}(\mu_{1,s}-\mu_{2,s})$.
Importantly, it is {\it spin density} and not {\it spin current} that is
dual to the charge current, which is a direct consequence
of the duality relation (\ref{eq:duality}). Physically,
$V_c$ and $V_s$ can be easily generated by two different voltage configurations of the four terminal setup,
as shown in Fig.~\ref{fig:duality}.
In these two symmetrical bias configurations, we find either charge current or spin density but no spin current. This is different for unsymmetrical bias configurations where charge current and spin density are usually accompanied by spin current as well.


Up to now, we have discussed transport properties of two hTLLs and shown
a simple relation between charge current and spin density
in two (dual) voltage configurations. In the following, we would like to
take into account a junction structure introducing scattering between
two hTLLs within a region of finite length $d$. We consider two different scenarios: (i) the short junction (SJ) with
$d\ll\lambda_f$ and (ii) the long junction (LJ) with $d\gg\lambda_f$, where $\lambda_f$ is the electron Fermi wave length.
For SJ, we can neglect the length of the scattering region and model it as a quantum point contact, while for LJ, we can regard
the scattering region as a sTLL with finite length.

\begin{figure}[h]
\begin{center}
\includegraphics[width=3in,angle=0]{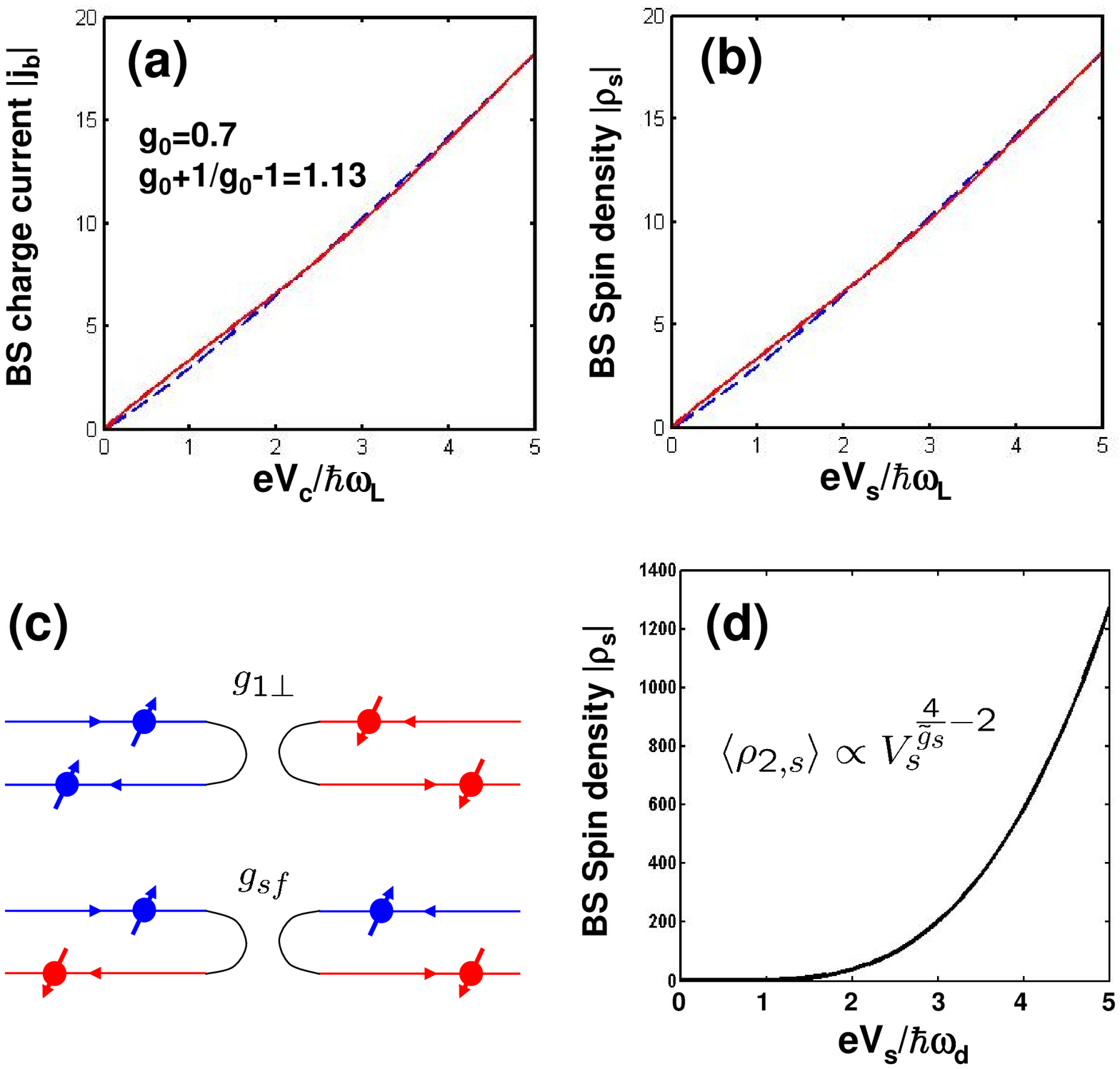}
\end{center}
\caption{ (Color online) (a) The backscattering current as a function of charge bias $V_c$ generated by the spin conserved $t_c$-term
for the voltage configuration of Fig.~\ref{fig:duality}(a).
The blue dashed line corresponds to the charge current $\langle j^{(0)}_{2,c}\rangle$,
while the red solid line additionally includes the finite length corrections.
The current unit is $\frac{2et^2_c\tau_{\rm cu}^{g_0+\frac{1}{g_0}}}
{\hbar^2\omega_L}$ . (b) is similar to (a), except that the spin density $\langle \rho^{(0)}_{2,s}\rangle$ and $\langle\rho_{2,s}\rangle$ as a function of the spin bias $V_s$
are generated by the spin-flip $t_s$-term for the voltage configuration of Fig.~\ref{fig:duality}(b). The spin density unit is $\frac{2t^2_s\tau_{\rm cu}^{g+\frac{1}{g}}}{\hbar^2 v_f\omega_L}$. 
(c) Possible two-particle backscattering terms:
spin conserved backscattering $g_{1\perp}$ and spin-flip backscattering $g_{sf}$.
(d) The voltage dependence of the correction to the spin density $\langle \rho_{2,s}\rangle$
due to the two-particle spin-flip term $g_{sf}$. The spin density unit is $\frac{(g_{sf}d)^2\tilde{g}_s(\tau_{\rm cu})^{\frac{4}{\tilde{g}_s}}}{2(\hbar\omega_d)^2d}$.
In all expressions above, we use $\omega_L=\frac{v_f}{g_0L}$, $\omega_d=\frac{v_f}{g_{s}d}$, $g_0=0.7$
and $g_{s}=1$.}
\label{fig:current}
\end{figure}

{\it Short junction case.--} In the following, we concentrate on the experimentally relevant regime $1/\sqrt{3}<g_0<1$ \cite{teo2009}. Then, all possible one-particle and two-particle scattering terms will be irrelevant \cite{teo2009}.
Hence, we can safely treat the scattering Hamiltonian $\hat{H}_t$ as a perturbation.
For the SJ case, two types of one particle scattering terms \cite{teo2009},
which preserve time reversal symmetry, are taken into account
(see Fig.~\ref{fig:configuration}(b)),
given by $\hat{H}_t=i t_c \sum_{\sigma=\pm}\eta^{R\sigma}\eta^{L\sigma}
    \sin(\sqrt{\pi}\Xi_\sigma)
    +it_s\sum_{a=\pm}\eta^{a\uparrow}\eta^{a\downarrow}\sin(\sqrt{\pi}\Psi_a)$
    where $t_c$ term preserves spin and $t_s$ term flips spin.
Here,   $\Xi_\sigma=\sqrt{2}(\varphi_c+\sigma\varphi_s)$, $\Psi_a=\sqrt{2}(\varphi_s+a\theta_s)$, and
 $\eta^{a\sigma}$ is the so-called Klein factor.
We now perform the perturbative calculation of our four terminal system
within the non-equilibrium Keldysh formalism \cite{kleinert1995,dolcini2005}.
All physical quantities can be related to expectation values
of boson fields.
Treating $\hat{H}_t$ as a perturbation, we can expand any physical quantity $\hat{O}$
($\hat{O}=\hat{j}_c,\hat{\rho}_s$) in powers of $t_{c(s)}$, e.g.
$\langle\hat{O}\rangle\approx\langle\hat{O}_{0}\rangle+\langle\hat{O}_{2}\rangle$
up to second order.
Explicit expressions for $\hat{j}_c$ and $\hat{\rho}_s$
in terms of expectation values of boson fields are given in the Appendices.
For clarity, we further divide the operator $\hat{O}_{2}$ into two parts $\langle \hat{O}_{2}\rangle
=\langle \hat{O}_{2}^{(0)}\rangle+\langle \hat{O}_{2}^{(1)}\rangle$,
where $\langle \hat{O}_{2}^{(0)}\rangle$ is calculated on the basis of $L\rightarrow\infty$
correlation functions, while
$\langle \hat{O}_{2}^{(1)}\rangle$ contains all finite length corrections.
Then, analytical expressions for the charge current and spin
density are readily obtained and given by $\langle \hat{j}^{(0)}_{2,c}\rangle=
-\frac{e\pi t^2_c\tau_{\rm cu}^{g_0+\frac{1}{g_0}}}
	{\hbar^2\omega_L\Gamma\left( g_0+\frac{1}{g_0} \right)}
	{\rm sgn}\left( V_c\right)\left|\frac{eV_c}{\hbar\omega_L}\right|^{g_0+\frac{1}{g_0}-1}$
and $	\langle \hat{\rho}^{(0)}_{2,s}\rangle=\frac{\pi t^2_s\tau_{\rm cu}^{g_0+\frac{1}{g_0}}}
	{\hbar^2 v_f\omega_L\Gamma\left( g_0+\frac{1}{g_0} \right)}
	{\rm sgn}\left( V_s \right)\left|\frac{eV_s}{\hbar\omega_L}  \right|^{g_0+\frac{1}{g_0}-1}$,
respectively, where $\omega_L=\frac{v_f}{g_0L}$ and $\tau_{\rm cu}$ is the short time cutoff. 
We note that the spin conserved scattering $t_c$ can only couple
to $V_c$ and, hence, reduce the total charge current in the voltage configuration of
Fig.~\ref{fig:duality}(a),
while the spin-flip scattering $t_s$, coupling to $V_s$,
decrease the spin density in the configuration of Fig.~\ref{fig:duality}(b).
In the absence of finite length corrections, both charge current and spin density depend in a simple power law fashion on the applied voltage,
in agreement with earlier work based on a renormalization group analysis \cite{hou2009,teo2009,strom2009,tanaka2009}.
Furthermore, we find that the ratio between charge current and spin density in the two dual voltage configurations is
$\langle \hat{j}^{(0)}_{2,c}\rangle/\langle \hat{\rho}^{(0)}_{2,s}\rangle
=ev_f\frac{t^2_c}{t^2_s}$,
which can be used to obtain information about the scattering strength
for the different types of scattering. In the finite length case, we use
the numerical method of Ref.~[\onlinecite{dolcini2005}] to evaluate $\langle \hat{j}_{2,c}\rangle$ and $\langle \hat{\rho}_{2,s}\rangle$.
As shown in Fig.~\ref{fig:current}(a) and (b), the finite length will introduce oscillations
in both the backscattering current and the spin density, which originates from Fabry-Perot-type interferences
of plasmonic excitation \cite{safi1995}. However, since the power law $g_0+\frac{1}{g_0}-1$ is always larger
than 1 for positive $g_0$, $\langle \hat{j}^{(0)}_{2,c}\rangle$
($\langle \hat{\rho}^{(0)}_{2,s}\rangle$) will increase rapidly
with $V_c$ ($V_s$) and dominate the oscillatory corrections for large $V_c$ ($V_s$). We conclude that finite length corrections are not very important in this setup.

{\it Long junction case.--}
Now we consider the opposite limit $d\gg\lambda_f$
as shown in Fig.~\ref{fig:configuration}(c), which could be achieved by gradually narrowing the QSH sample
into a 1D wire experimentally. In the following, we analyze finite size effects related to $d$ and
assume that $L\gg d$, hence $L \rightarrow \infty$ is a reasonable approximation.
For simplicity, we model the LJ as a sTLL described by $\hat{H}_t=\hat{H}_{t0}+\hat{H}_{t2}$
with $\hat{H}_{t0}=\hat{H}^c_{t0}+\hat{H}^s_{t0}$ and
$\hat{H}^c_{t0}=\frac{\hbar v_f}{2}\int dx\left[ (\partial_x\theta_c)^2+\frac{1}{g_0^2}
    (\partial_x\varphi_c)^2\right]$,
   $\hat{H}^s_{t0}=\frac{\hbar v_f}{2}\int dx\left[ (\partial_x\theta_s)^2+
   \frac{1}{g_{s}^2} (\partial_x\varphi_s)^2\right]$
for $|x|<d/2$. Compared to Eq.~(\ref{eq:H0hl2}), we find that the charge sector remains unchanged at the step from hTLL to sTLL, but the spin sector shows a stepwise variation of the interaction parameter as well as the velocity of spin excitations. Note that for finite spin-orbit coupling $g_s$ can be renormalized away from its non-interacting value $g_s=1$, see \cite{gritsev2005,tanaka2009}.
Since the Hamiltonian in the charge sectors remains unchanged, the interaction between two hTLL (as described in $\hat{H}^c_{t0}$) will not affect the charge current $\langle\hat{j}_{c}\rangle$ at all. For the spin sector,
besides $\hat{H}^s_{t0}$ we consider two additional interaction terms \cite{gritsev2005,tanaka2009}
$\hat{H}_{t2}=\int^{\frac{d}{2}}_{-\frac{d}{2}}
	dx\left[ g_{1\perp}\cos\left( \sqrt{8\pi}\varphi_s \right)
	+g_{sf}\cos\left( \sqrt{8\pi}\theta_s \right)\right]$
where the $g_{1\perp}$-term is related to spin conserved backscattering
$\psi^{\dag}_{L,\uparrow}\psi^{\dag}_{R,\downarrow}\psi_{L,\downarrow}\psi_{R,\uparrow}
+h.c.$
while the $g_{sf}$-term is related to spin-flip backscattering
$\psi^{\dag}_{L,\downarrow}\psi^{\dag}_{R,\downarrow}\psi_{L,\uparrow}\psi_{R,\uparrow}
+h.c.$, see Fig.~\ref{fig:current}(c). These are the most important perturbations to the spin sector in the absence of impurity scattering. Since the $g_{1\perp}$-term conserves spin, it will not
influence the spin density. Thus, we focus on the $g_{sf}$-term below.
Up to second order perturbation theory, we obtain the following
correction to the spin density
$\langle \rho_{2,s}\rangle=-\frac{(g_{sf}d)^2\tilde{g}_s}{(\hbar\omega_d)^2d}\int^{1}_{0}dR
	\int^{1-|R|}_{0}dr f(r,R)$
	where $\tilde{g}_s=g_{s}g_0$ and the function $f(r,R)$ is specified
	in Eq.(\ref{eq:rhostheta1}) of App. B.
The above integration can be easily
evaluated numerically (similar to the Coulomb drag problem in \cite{peguiron2007})
and the obtained correction
to the spin density is shown in Fig.~\ref{fig:current}(d). To make analytical progress,
we can again divide the obtained spin density into two parts
$\langle \rho_{2,s}\rangle=\langle \rho^{(0)}_{2,s}\rangle+
\langle \rho^{(1)}_{2,s}\rangle$, with $\langle \rho^{(0)}_{2,s}\rangle$
for the infinite $d$ and $\langle \rho^{(1)}_{2,s}\rangle$ for the
finite length corrections. It is found that $\langle \rho^{(0)}_{2,s}\rangle$
always dominates over $\langle \rho^{(1)}_{2,s}\rangle$. In the limit
$eV_s/\hbar\omega_d\gg 1$, we obtain $\langle \rho^{(0)}_{2,s}\rangle\approx{\rm sgn}\left( eV_s \right)
\frac{(g_{sf}d)^2\pi^2\tilde{g}_s}{d(\hbar\omega_d)^2\Gamma^2\left(\frac{2}{\tilde{g}_s}\right)}\tau_{\rm cu}^{\frac{4}{\tilde{g}_s}}
\left|\frac{eV_s}{\hbar\omega_d}\right|^{\frac{4}{\tilde{g}_s}-2}$
with $\omega_d=\frac{v_f}{g_{s}d}$.
\footnote{We mention here the validity regime of our perturbation theory. Note that for $g_0<1$ and $g_{s}\approx 1$, the power law exponent $\frac{4}{\tilde{g}_s}-2$
will always be larger than 2. Thus, the second order correction to the spin density $\langle \rho_{2,s}\rangle$ will increase faster than the zero order term $\langle \rho_{0,s}\rangle$ as a function of bias voltage. Hence, our perturbation theory is only justified as long as
$\tau_{\rm cu}^{4/\tilde{g}_s}\left( \frac{g_{sf}d}{\hbar\omega_d} \right)^2
\left| \frac{eV_s}{\hbar\omega_d} \right|^{\frac{4}{\tilde{g}_s}-3}\ll1$.}



{\it Conclusions.--}
We have analyzed the charge current and spin density in a four-terminal setup based on two hTLL coupled to non-interacting electron reservoirs. Different types of scattering mechanisms between the edges are taken into account, particularly short junctions and long junctions. It is shown that different power law dependencies as a function of bias voltages applied to the four terminals can be used to distinguish the scattering mechanisms. A simple duality relation between charge current and spin density has been discovered. Remarkably, all spin-related observables can be measured by straightforward charge measurements in the four-terminal configuration. It is interesting to ask the question whether the spin density in such a setup can also be measured by other means (e.g. as a test of the model). Taking typical values for $\hbar v_f=3{\rm eV}\cdot$\AA $\,$ and $eV_s=5{\rm meV}$, we find that the zero order spin density is about $\langle \hat{\rho}_{0,s}\rangle\approx 5.3 {\rm \mu m}^{-1}$. This may be detected by state-of-the-art local Faraday/Kerr rotation \cite{kato2004}.

We would like to thank the Humboldt foundation (CXL), the Emmy-Noether program (PR), and the DFG-JST Research Unit ``Topological electronics''
(JCB and BT) for funding as well as K. Le Hur, L.W. Molenkamp, J. Maciejko and S.-C. Zhang for interesting discussions.

\appendix
\section{Non-equilibrium Keldysh formalism of the spin-charge transport}
In this Appendix, we want to establish the general formalism used for the calculation of spin-charge transport properties
of our setup. The generating function of the system which is the starting point for all our calculations is most conveniently written down in
the helical basis. However, to make contact with the notation used in large parts of the main text, its representation in the spin-charge basis will be presented later on using Eq. (1) of the main text to express the non-chiral boson fields in terms of spin-charge fields.
For our non-equilibrium transport calculations we use the Keldysh formalism\cite{kleinert1995,dolcini2005},
in which all the field operators have a time variable residing on the Keldysh contour. This contour consists of the upper (forward)
and the lower (backward) branch. In real time representation, this time dependence enters the formalism via a so called Keldysh index $\eta$, where $\varphi^\eta$ and $\theta^\eta$
denote the boson and the dual boson field on the upper branch ($\eta=+$) and the lower branch ($\eta=-$), respectively.
In the helical basis, the generating function is then given by
\begin{eqnarray}
	&&Z[J]=\int \frac{\Pi_{\alpha=\pm}\mathcal{D}\varphi^{\pm}_\alpha\mathcal{D}\theta_\alpha^{\pm}}{N_Z}
	\mbox{exp}\left\{ i\sum_{\alpha=\pm}S^\alpha_0[\varphi_\alpha^\pm,\theta_\alpha^\pm]\right\}\nonumber\\
	&&    \mbox{exp}\left\{i\int_{-\infty}^{\infty} dt\sum_{\eta=\pm}
    \left[-\eta \left( H^\eta_V[\varphi^\eta_\pm,\theta^\eta_\pm]+H^\eta_t[\varphi^\eta_\pm,\theta^\eta_\pm] \right)
    \right.\right.\nonumber\\
    &&\left.\left. +\sum_{\alpha=\pm}\left( \int dx J_\alpha^\varphi(\bold{r})
    \varphi^\eta_\alpha(\bold{r})+\int dx J_\alpha^\theta(\bold{r})
    \theta^\eta_\alpha(\bold{r})  \right)\right]\right\},\nonumber\\\label{eq:Z1}
\end{eqnarray}
where $S^\alpha_0$ is the action of the free Boson field at the edge $\alpha=\pm$,
which reads \cite{giamarchi2004}
\begin{eqnarray}
	&&S^\alpha_0=\sum_{\eta=\pm}\eta\int dt\int dx\left[ -\partial_t\varphi^\eta_\alpha\partial_x\theta^\eta_\alpha
	-\frac{v_f}{2}\left( \partial_x\theta^\eta_\alpha \right)^2\right.\nonumber\\
	&&\left.-\frac{v_f}{2g(x)^2}\left( \partial_x\varphi^\eta_\alpha \right)^2 \right].
	\label{eq:S01}
\end{eqnarray}
Here we have set $\hbar=1$. $H_V$ which models the coupling to the leads can be written as\cite{dolcini2005}
\begin{eqnarray}
	\hat{H}^\eta_V=\int dt \sum_{\alpha}\left[\int \frac{dx}{\sqrt{\pi}}
    \partial_x\mu^\varphi_\alpha(x)\varphi_\alpha^\eta+\int \frac{dx}{\sqrt{\pi}}
    \partial_x\mu^\theta_\alpha(x)\theta_\alpha^\eta\right],\nonumber\\
	\label{eq:Hv1}
\end{eqnarray}
where $\mu_\alpha^\varphi$ and $\mu_\alpha^\theta$ are the generalized chemical potentials coupling the helical liquids
with the leads. Usually $\mu_\alpha^\varphi\neq0$ and $\mu_\alpha^\theta=0$ for an electric chemical
potential. However, to keep the whole formalism as general as possible, we consider both contributions.
$H_t$ describes inter edge coupling terms giving rise to different scattering processes. The concrete form of $H_t$ depends
on the detailed mechanism of the scattering process under investigation and will be discussed below.
Throughout this work we treat $H_t$ as a perturbation and present the general formalism
up to the second order. The source terms $J_\alpha^\varphi
\varphi^\eta_\alpha$ and $J_\alpha^\theta\theta^\eta_\alpha$ introduced in $Z[J]$~ allow for the calculation of arbitrary correlation functions of interest by evaluating suitable functional derivatives of the generating function with respect to $J$~ at $J=0$.

Of course all calculations can equivalently be performed in the helical basis as well as in the spin-charge
basis. However, we first transform the full Hamiltonian into the spin-charge basis since most of the quantities we want to compute are naturally written in terms of spin and charge fields.
In the spin-charge basis, the generating function can be written as
\begin{eqnarray}
	&&Z[J]=\int \frac{\Pi_{\lambda=c,s}\mathcal{D}\varphi^{\pm}_\lambda\mathcal{D}\theta_\lambda^{\pm}}{N_Z}
	\mbox{exp}\left\{ i\sum_{\lambda=c,s}S^{\lambda}_0[\varphi_\lambda^\pm,\theta_\lambda^\pm]\right\}\nonumber\\
	&&   \mbox{exp}\left\{i\int_{-\infty}^{\infty} dt\sum_{\eta=\pm}
    \left[-\eta \left( H^\eta_V[\varphi^\eta_{c(s)},\theta^\eta_{c(s)}]+H^\eta_t[\varphi^\eta_{c(s)},\theta^\eta_{c(s)}] \right)
    \right.\right.\nonumber\\
    &&\left.\left. +\sum_{\lambda=c,s}\left( \int dx J_\lambda^\varphi(\bold{r})
    \varphi^\eta_\lambda(\bold{r})+\int dx J_\lambda^\theta(\bold{r})
    \theta^\eta_\lambda(\bold{r})  \right)\right]\right\}
    \label{eq:Z2}
\end{eqnarray}
with
\begin{eqnarray}
	&&S^c_0=\sum_{\eta=\pm}\eta\int dt\int dx\left[ -\partial_t\varphi^\eta_c\partial_x\theta^\eta_c
	-\frac{v_f}{2}\left( \partial_x\theta^\eta_c \right)^2\right.\nonumber\\
	&&\left.-\frac{v_f}{2g(x)^2}\left( \partial_x\varphi^\eta_c \right)^2 \right],\nonumber\\
        &&S^s_0=\sum_{\eta=\pm}\eta\int dt\int dx\left[ -\partial_t\theta^\eta_s\partial_x\varphi^\eta_s
	-\frac{v_f}{2}\left( \partial_x\varphi^\eta_s \right)^2\right.\nonumber\\
	&&\left.-\frac{v_f}{2g(x)^2}\left( \partial_x\theta^\eta_s \right)^2 \right],
	\label{eq:S02}
\end{eqnarray}
and
\begin{eqnarray}
	\hat{H}^\eta_V=\sum_{\lambda}\int \frac{dtdx}{\sqrt{2\pi}}  \left[\partial_x\mu^\varphi_\lambda(x)
	\varphi_\lambda^\eta+\partial_x\mu^\theta_\lambda(x)\theta_\lambda^\eta\right],
	\label{eq:Hv2}
\end{eqnarray}
where $\mu^\varphi_c=\mu^\varphi_++\mu^\varphi_-$, $\mu^\theta_s=\mu^\varphi_+-\mu^\varphi_-$,
$\mu^\theta_c=\mu^\theta_++\mu^\theta_-$ and $\mu^\varphi_s=\mu^\theta_+-\mu^\theta_-$.
Now let us introduce the following notations
\begin{eqnarray}
	&&\Phi_\lambda=\left(
	\begin{array}{c}
		\varphi^+_\lambda(\bold{r})\\
		\varphi^-_\lambda(\bold{r})\\
		\theta^+_\lambda(\bold{r})\\
		\theta^-_\lambda(\bold{r})
	\end{array}
	\right),\\
	&&\bold{J}_\lambda=\left(
        \begin{array}{c}
        -\sqrt{\frac{1}{\pi}}\partial_x\mu^\varphi_\lambda(\bold{r})\\
        \sqrt{2}J^\varphi_\lambda(\bold{r})\\
        -\sqrt{\frac{1}{\pi}}\partial_x\mu^\theta_\lambda(\bold{r})\\
        \sqrt{2}J^\theta_\lambda(\bold{r})\\
        \end{array}
        \right)\\
	&&Q=\frac{1}{\sqrt{2}}\delta(\bold{r}-\bold{r}')\left(
	\begin{array}{cccc}
		1&-1&0&0\\
		1&1&0&0\\
		0&0&1&-1\\
		0&0&1&1
	\end{array}
	\right)\\
	&&\bold{C}_{\lambda}=\left(
	\begin{array}{cccc}
	\mathcal{C}^{++}_{\lambda}(\bold{r},\bold{r}')&\mathcal{C}^{+-}_{\lambda}(\bold{r},\bold{r}')&
	\mathcal{F}^{++}_{\lambda}(\bold{r},\bold{r}')&\mathcal{F}^{+-}_{\lambda}(\bold{r},\bold{r}')\\
	\mathcal{C}^{-+}_{\lambda}(\bold{r},\bold{r}')&\mathcal{C}^{--}_{\lambda}(\bold{r},\bold{r}')&
        \mathcal{F}^{-+}_{\lambda}(\bold{r},\bold{r}')&\mathcal{F}^{--}_{\lambda}(\bold{r},\bold{r}')\\
	\mathcal{Q}^{++}_{\lambda}(\bold{r},\bold{r}')&\mathcal{Q}^{+-}_{\lambda}(\bold{r},\bold{r}')&
	\mathcal{D}^{++}_{\lambda}(\bold{r},\bold{r}')&\mathcal{D}^{+-}_{\lambda}(\bold{r},\bold{r}')\\
	\mathcal{Q}^{-+}_{\lambda}(\bold{r},\bold{r}')&\mathcal{Q}^{--}_{\lambda}(\bold{r},\bold{r}')&
        \mathcal{D}^{-+}_{\lambda}(\bold{r},\bold{r}')&\mathcal{D}^{--}_{\lambda}(\bold{r},\bold{r}')
	\end{array}
	\right)\nonumber\\\label{eq:Correlation_def1}
\end{eqnarray}
where $\mathcal{C}^{\eta\eta'}_{\lambda}(\bold{r},\bold{r}')=\langle \varphi_{\lambda}^\eta(\bold{r})
\varphi_{\lambda}^{\eta'}(\bold{r}')\rangle_0$, $\mathcal{D}^{\eta\eta'}_{\lambda}(\bold{r},\bold{r}')
=\langle \theta_{\lambda}^\eta(\bold{r})\theta_{\lambda}^{\eta'}(\bold{r}')\rangle_0$,
$\mathcal{F}^{\eta\eta'}_{\lambda}(\bold{r},\bold{r}')=\langle \varphi_{\lambda}^\eta(\bold{r})
\theta_{\lambda}^{\eta'}(\bold{r}')\rangle_0$, and $\mathcal{Q}^{\eta\eta'}_{\lambda}(\bold{r},\bold{r}')
=\langle \theta_{\lambda}^\eta(\bold{r})\varphi_{\lambda}^{\eta'}(\bold{r}')\rangle_0
=(\mathcal{F}^{\eta',\eta}_{\lambda}(\bold{r}',\bold{r}))^\dag$.
With the compact definitions above, it is easy to check that the generating function can be written as
\begin{eqnarray}
	&&Z[J]=\int \frac{\Pi_{\lambda=c,s}\mathcal{D}\Phi_\lambda}{N_Z}
	e^{\sum_\lambda(-\frac{1}{2}\Phi_\lambda^T\bold{C}_{\lambda}^{-1}\Phi_\lambda
	+i\bold{J}_\lambda^TQ\Phi_\lambda)}\nonumber\\
	&&e^{-i\sum_\eta\eta\int^{\infty}_{-\infty}dtH_t[\Phi_c,\Phi_s]}.
	\label{eq:Z3}
\end{eqnarray}
Here the superscript T represents the matrix transpose. Note that the matrix product in the above equations includes an integration over space and time. In order to separate the free part of the generating function from the terms to be treated perturbatively, we apply the following shift to the vector of boson fields
\begin{eqnarray}
	\tilde{\Phi}_\lambda=\Phi_\lambda-\bold{A}_\lambda[J_\lambda], \qquad
	\bold{A}_\lambda[J_\lambda]=i\bold{C}_{\lambda}Q^T\bold{J}_\lambda,
	\label{eq:AJ1}
\end{eqnarray}
and explicitly ${\bf A}_\lambda[J_\lambda]=[A_\lambda^{+,\varphi},A_\lambda^{-,\varphi},
A_\lambda^{+,\theta},A_\lambda^{-,\theta}]^T$ with
\begin{eqnarray}
	&&A_\lambda^{\eta,\varphi}[J_\lambda]=-\frac{i}{\sqrt{2\pi}}\int d\bold{r}'\left(\mathcal{C}^R_{\lambda}
	(\bold{r},\bold{r}')\partial_x\mu^\varphi_{\lambda}(\bold{r}')\right.\nonumber\\
	&&\left.+\mathcal{F}^R_\lambda
    (\bold{r},\bold{r}')\partial_x\mu^\theta_\lambda(\bold{r}')\right)+i\int d\bold{r}'
    (\mathcal{C}_\lambda^K(\bold{r},\bold{r}')+\eta\mathcal{C}_\lambda^A(\bold{r},\bold{r}'))
    J^\varphi_\lambda(\bold{r}')\nonumber\\
    &&+i\int d\bold{r}'(\mathcal{F}_\lambda^K(\bold{r},\bold{r}')+
    \eta\mathcal{F}_\lambda^A(\bold{r},\bold{r}'))J^\theta_\lambda(\bold{r}')\\
    &&A_\lambda^{\eta,\theta}[J_\lambda]=-\frac{i}{\sqrt{2\pi}}\int d\bold{r}'\left(\mathcal{Q}^R_\lambda
    (\bold{r},\bold{r}')\partial_x\mu^\varphi_\lambda(\bold{r}')\right.\nonumber\\
    &&\left.+\mathcal{D}^R_\lambda
    (\bold{r},\bold{r}')\partial_x\mu^\theta_\lambda(\bold{r}')\right)+i\int d\bold{r}'
    (\mathcal{Q}_\lambda^K(\bold{r},\bold{r}')+\eta\mathcal{Q}_\lambda^A(\bold{r},\bold{r}'))
    J^\varphi_\lambda(\bold{r}')\nonumber\\
        &&+i\int d\bold{r}'(\mathcal{D}_\lambda^K(\bold{r},\bold{r}')+
    \eta\mathcal{D}_\lambda^A(\bold{r},\bold{r}'))J^\theta_\lambda(\bold{r}').
    \label{eq:GFD_A}
\end{eqnarray}
The generating function then factorizes as follows
\begin{eqnarray}
	Z[J]=Z_{0,c}[J_c]Z_{0,s}[J_s]Z_t[J]
	\label{eq:Zdec1}
\end{eqnarray}
where
\begin{eqnarray}
	&&Z_{0,\lambda}[J_\lambda]=e^{-\frac{1}{2}\bold{J}^T_\lambda\tilde{\bold{C}}_{\lambda}\bold{J}_\lambda},\\
	&&\tilde{\bold{C}}_{\lambda}=Q\bold{C}_{\lambda}Q^T=\nonumber\\
	&&\left(\begin{array}{cccc}
		0&\mathcal{C}^A_{\lambda}(\bold{r},\bold{r}')&0&
		\mathcal{F}^A_{\lambda}(\bold{r},\bold{r}')\\
		\mathcal{C}^R_{\lambda}(\bold{r},\bold{r}')&
		\mathcal{C}^K_{\lambda}(\bold{r},\bold{r}')&
		\mathcal{F}^R_{\lambda}(\bold{r},\bold{r}')&
		\mathcal{F}^K_{\lambda}(\bold{r},\bold{r}')\\
		0&\mathcal{Q}^A_{\lambda}(\bold{r},\bold{r}')&0&
		\mathcal{D}^A_{\lambda}(\bold{r},\bold{r}')\\
		\mathcal{Q}^R_{\lambda}(\bold{r},\bold{r}')&
		\mathcal{Q}^K_{\lambda}(\bold{r},\bold{r}')&
		\mathcal{D}^R_{\lambda}(\bold{r},\bold{r}')&
		\mathcal{D}^K_{\lambda}(\bold{r},\bold{r}')
	\end{array}
	\right)\nonumber\\
	\label{eq:Zdec2}
\end{eqnarray}
with $\mathcal{C}^R_{\lambda}=\theta(t-t')\langle [\varphi_\lambda(\bold{r}),
\varphi_\lambda(\bold{r}')]\rangle_0$, $\mathcal{C}^A_{\lambda}(\bold{r},\bold{r}')
=-\theta(t'-t)\langle [\varphi_\lambda(\bold{r}),\varphi_\lambda(\bold{r}')]\rangle_0$
and $\mathcal{C}^K_{\lambda}=\langle\{\varphi_\lambda(\bold{r}),\varphi_\lambda(\bold{r}')\}\rangle_0$,
and similar definitions for the other correlation functions $\mathcal{D}$, $\mathcal{F}$
and $\mathcal{Q}$. $Z_t[J]$ is given by
\begin{eqnarray}
	&&Z_t[J]=\int \frac{\Pi_{\lambda=c,s}\mathcal{D}\tilde{\Phi}_\lambda}{N_Z}
	e^{-\frac{1}{2}\sum_\lambda\tilde{\Phi}^T_\lambda\bold{C}_{\lambda}^{-1}
	\tilde{\Phi}_\lambda}\nonumber\\
	&&e^{-i\sum_\eta\eta\int^{\infty}_{-\infty}dt
	H_t[\tilde{\Phi}_c+{\bf A}_c,\tilde{\Phi}_s+{\bf A}_s]}\nonumber\\
	&&=\langle e^{-i\sum_\eta\eta\int^{\infty}_{-\infty}dt
	H_t[\tilde{\Phi}_c+{\bf A}_c,\tilde{\Phi}_s+{\bf A}_s]}\rangle_0.
	\label{eq:Z21}
\end{eqnarray}
Next we need to relate the physical quantities to the generating
function. The density is defined as
$\langle\hat{\rho}_\lambda\rangle=\sqrt{\frac{2}{\pi}}\partial_x\langle\varphi_\lambda\rangle$
and the current as
$\langle\hat{j}_\lambda\rangle=e\sqrt{\frac{2}{\pi}}\partial_t\langle\varphi_\lambda\rangle=-ev_f\sqrt{\frac{2}{\pi}}\partial_x\langle\theta_\lambda\rangle$.
As already mentioned, expectation values like $\langle\varphi_\lambda\rangle$ and $\langle\theta_\lambda\rangle$ can be conveniently calculated
from the functional derivatives of the generating function $Z[J]$ with respect to $J^\varphi_\lambda$
and $J^\theta_\lambda$. A direct calculation shows that
\begin{eqnarray}
	&&\langle \varphi_\lambda(\bold{r})\rangle=\frac{1}{2}\sum_\eta \langle \varphi^\eta_\lambda(\bold{r})\rangle
	=-\frac{i}{2}\frac{1}{Z[0]}\left.\frac{\delta Z[J]}{\delta J^\varphi_\lambda(\bold{r})}\right|_{J=0}\nonumber\\
	&&=-\frac{i}{2}\left.\left[ \frac{\delta Z_{0,\lambda}[J_\lambda]}
	{\delta J^\varphi_\lambda(\bold{r})}+\frac{1}{Z_t[0]}\frac{\delta Z_t[J]}
	{\delta J^\varphi_\lambda(\bold{r})} \right]\right|_{J=0},
    \label{eq:varphi1}
\end{eqnarray}
with a similar expression holding for $\langle\theta_\lambda\rangle$.
From (\ref{eq:varphi1}), we find that $\langle\varphi_\lambda\rangle$ can be decomposed
into two parts: one is the zero order term coming from $Z_{0,\lambda}[J]$, and
the other one is the scattering term coming from $Z_t[J]$. Consequently, any physical
quantity $\hat{O}$ can also be divided into two parts
$\langle\hat{O}\rangle=\langle\hat{O}_{0}\rangle+\langle\hat{O}_{2}\rangle$. After a lengthy derivation, the density and the current can be expressed as
\begin{eqnarray}
    &&\langle \rho_\lambda(\bold{r})\rangle=\sqrt{\frac{2}{\pi}}\partial_x\langle\varphi_\lambda\rangle
    =\langle \rho_{0,\lambda}(\bold{r})\rangle+\langle \rho_{2,\lambda}(\bold{r})\rangle,\label{eq:rho_1}\\
    &&\langle \rho_{0,\lambda}(\bold{r})\rangle=-\frac{i}{\pi}\int
    d\bold{r}'\partial_x\left(
    \begin{array}{cc}
	    \mathcal{C}^R_{\lambda}(\bold{r},\bold{r}')&
	\mathcal{F}^R_{\lambda}(\bold{r},\bold{r}')
    \end{array}
    \right)\nonumber\\
    &&\left(
    \begin{array}{c}
	    \partial_{x'}\mu^\varphi_\lambda(\bold{r}')\\
	    \partial_{x'}\mu^\theta_\lambda(\bold{r}')
    \end{array}
    \right),\label{eq:rho_2}\\
    &&\langle \rho_{2,\lambda}(\bold{r})\rangle=i\sqrt{\frac{2}{\pi}}\int d\bold{r}'\partial_x
    \mathcal{C}^R_{\lambda}(\bold{r},\bold{r}')\left\langle
    \hat{f}^{\varphi}_{\lambda}(\bold{r}')\right\rangle_t\nonumber\\
    &&+i\sqrt{\frac{2}{\pi}}\int d\bold{r}'\partial_x\mathcal{F}^R_{\lambda}
    (\bold{r},\bold{r}')\left\langle\hat{f}^{\theta}_{\lambda}(\bold{r}')\right\rangle_t,
    \label{eq:rho_3}
\end{eqnarray}
\begin{eqnarray}
    &&\langle j_\lambda(\bold{r})\rangle=-ev_f\sqrt{\frac{2}{\pi}}\partial_x\langle\theta_\lambda\rangle
    =\langle j_{0,\lambda}(\bold{r})\rangle+\langle j_{2,\lambda}(\bold{r})\rangle,\nonumber\\
    \label{eq:current_1}\\
    &&\langle j_{0,\lambda}(\bold{r})\rangle=\frac{iev_f}{\pi}\int
    d\bold{r}'\partial_x\left(
    \begin{array}{cc}
	    \mathcal{Q}^R_{\lambda}(\bold{r},\bold{r}')&
	    \mathcal{D}^R_{\lambda}(\bold{r},\bold{r}')
    \end{array}
    \right)\nonumber\\
    &&\left(
    \begin{array}{c}
	    \partial_{x'}\mu^\varphi_\lambda(\bold{r}')\\
	    \partial_{x'}\mu^\theta_\lambda(\bold{r}')
    \end{array}
    \right),\label{eq:current_2}\\
        &&\langle j_{2,\lambda}(\bold{r})\rangle=-iev_f\sqrt{\frac{2}{\pi}}\int d\bold{r}'\partial_x
    \mathcal{Q}^R_{\lambda}(\bold{r},\bold{r}')\left\langle
    \hat{f}^{\varphi}_{\lambda}(\bold{r}')\right\rangle_t\nonumber\\
    &&-iev_f\sqrt{\frac{2}{\pi}}\int d\bold{r}'\partial_x\mathcal{D}^R_{\lambda}(\bold{r},\bold{r}')
    \left\langle\hat{f}^{\theta}_{\lambda}(\bold{r}')\right\rangle_t
    \label{eq:current_3}
\end{eqnarray}
where $\hat{f}^{\varphi}_{\lambda}=-\frac{\delta \hat{H}_2[\Phi+{\bf A}]}{\delta \varphi^+_\lambda}$
and $\hat{f}^{\theta}_{\lambda}=-\frac{\delta \hat{H}_2[\Phi+{\bf A}]}{\delta \theta^+_\lambda}$,
and $\langle \cdots \rangle_t=\frac{1}{Z_t[0]}\int \frac{\Pi_\lambda\mathcal{D}_\lambda\Phi}{N_Z}
\cdots e^{-\frac{1}{2}\sum_\lambda\Phi^T\bold{C}_\lambda^{-1}\Phi_\lambda}
e^{-i\sum_\eta\eta\int dtH_2[\Phi+{\bf A}]}$
 means the average along the Keldysh contour with respect to the
action $\sum_\lambda S_{0,\lambda}-\int dt H_t[\Phi_c+{\bf A}_c,\Phi_s+{\bf A}_s]$.
Here we use $\Phi$ instead of $\tilde\Phi$ to keep our notation simple.
The expressions (\ref{eq:rho_1}) to (\ref{eq:current_3}) are the central results for our
approach to the calculation of charge current and spin density.

\section{Charge current and Spin density}
In this section, we calculate the charge current and spin density in our model. Now we
consider the realistic electric chemical potential ($\mu^\varphi_\pm\neq0$
and $\mu^\theta_\pm=0$) so that only $\mu^\varphi_c$ and $\mu^\theta_s$ do not vanish.
Let us choose the form of $\mu$ in each edge as
$\partial_x\mu^\varphi_\pm=-\mu_{1,\pm}\delta(x+L/2)+\mu_{2,\pm}\delta(x-L/2)$,
then we have $\partial_x\mu^\varphi_c=-\mu_{1,c}\delta(x+L/2)+\mu_{2,c}\delta(x-L/2)$
with $\mu_{1(2),c}=\mu_{1(2),+}+\mu_{1(2),-} $ and
$\partial_x\mu^\theta_s=-\mu_{1,s}\delta(x+L/2)+\mu_{2,s}\delta(x-L/2)$
where $\mu_{1(2),s}=\mu_{1(2),+}-\mu_{1(2),-}$.
Moreover we define the voltage on the upper or lower edge as
$eV_{\alpha}=\mu_{1,\alpha}-\mu_{2,\alpha}$ and correspondingly
$eV_c=e\left( V_++V_- \right)/2=\left( \mu_{1,c}-\mu_{2,c} \right)/2$
and $eV_s=e\left( V_+-V_- \right)/2=\left( \mu_{1,s}-\mu_{2,s} \right)/2$.

From the expressions (\ref{eq:rho_2}) and (\ref{eq:current_2}),the zeroth order
of the charge current and spin density are easy to obtain, given the free correlation functions which will be discussed in some detail in the next section. The unperturbed contributions yield
\begin{eqnarray}
	\langle j_{0,c}\rangle=\frac{2e^2}{h}V_c,\qquad
	\langle \rho_{0,s}\rangle=-\frac{2e}{hv_f}V_s.
	\label{eq:zerojrho}
\end{eqnarray}
where we have recovered $\hbar$.

The influence of the scattering depends on its detailed mechanism. Here we discuss two opposite cases separately, the short junction and the
long junction. For the short junction, the scattering Hamiltonian
is given by
\begin{eqnarray}
	&&\hat{H}_t=\int dx \delta(x-x_0)\nonumber\\
	&&\left[ \tilde{t}_c(\psi^{\dag}_{R\uparrow}(x)\psi_{L\uparrow}(x)+\psi^{\dag}_{L\downarrow}(x)\psi_{R\downarrow}(x)
	+h.c.)\right.\nonumber\\
	&&\left.+\tilde{t}_s(\psi^{\dag}_{R\uparrow}(x)\psi_{R\downarrow}(x)-\psi^{\dag}_{L\downarrow}(x)\psi_{L\uparrow}(x)+h.c.) \right].
    \label{eq:MA_Htunnel}
\end{eqnarray}
In the bosonization form, we have
\begin{eqnarray}
    &&\hat{H}_t=\int dx \delta (x-x_0)\left[ it_c\sum_\sigma\eta^{R\sigma}\eta^{L\sigma}
    \sin(\sqrt{\pi}\Xi_\sigma)\right. \nonumber\\
    &&\left.+it_s\sum_a\eta^{a\uparrow}\eta^{a\downarrow}\sin(\sqrt{\pi}\Psi_a) \right]
    \label{eq:Hamt}
\end{eqnarray}
with $t_{c(s)}=\frac{\tilde{t}_{c(s)}}{\pi}$, $\Xi_\sigma=\sqrt{2}(\varphi_c+\sigma\varphi_s)$ and $\Psi_a=\sqrt{2}(\varphi_s+a\theta_s)$.
Given this concrete form of $\hat{H}_t$, the operators $\hat{f}^\varphi_\lambda$ and $\hat{f}^\theta_\lambda$ read
\begin{eqnarray}
    &&\hat{f}^{\varphi}_{c}(x')=-it_c\sqrt{2\pi}\delta(x'-x_0)\sum_\sigma\eta^{R\sigma}\eta^{L\sigma}\nonumber\\
    &&\cos\sqrt{2\pi}
    (\bar\varphi^+_c+\sigma\bar\varphi^+_s)\nonumber\\
        &&\hat{f}^{\theta}_{c}(x')=0\nonumber\\
    &&\hat{f}^{\varphi}_{s}(x')=-i\sqrt{2\pi}\delta(x'-x_0)\nonumber\\
    &&\left[ t_c\sum_\sigma\eta^{R\sigma}\eta^{L\sigma}\sigma\cos\sqrt{2\pi}(\bar\varphi^+_c+\right.\nonumber\\
    &&\left.\sigma\bar\varphi^+_s)+t_s\sum_a\eta^{a\uparrow}\eta^{a\downarrow}\cos\sqrt{2\pi}\left(
    \bar\varphi^+_s+a\bar\theta^+_s\right) \right]\nonumber\\
        &&\hat{f}^{\theta}_{s}(x')=-it_s\sqrt{2\pi}\delta(x'-x_0)\sum_a\eta^{a\uparrow}\eta^{a\downarrow}a\nonumber\\
	&&\cos\sqrt{2\pi}
    (\bar\varphi^+_s+a\bar\theta^+_s)
    \label{eq:currentoperator}
\end{eqnarray}
where $\bar\varphi^\eta_\lambda=\varphi^\eta_\lambda+A^{\eta,\varphi}_\lambda[0]$
and $\bar\theta^\eta_\lambda=\theta^\eta_\lambda+A^{\eta,\theta}_\lambda[0]$.
The average $\langle \hat{f}_\lambda\rangle_t=\langle \hat{f}_\lambda
e^{-i\sum_\eta\eta\int dtH_2[\Phi+{\bf A}]}\rangle_0$,
is then given by
\begin{eqnarray}
	&&\langle f^\varphi_c(x')\rangle_t=\frac{t^2_c\sqrt{2\pi}}{4}\delta(x'-x_0)
	\int d{\bf r}_1\delta(x_1-x_0)\nonumber\\
	&&\sum_{m,\sigma} me^{im\sqrt{\pi}\left( A^\Xi_\sigma({\bf r}')
	-A^{\Xi}_\sigma({\bf r}_1)\right)}\nonumber\\
	&&\left[ -\left\langle e^{i\sqrt{\pi}\Xi_\sigma({\bf r}')}
	e^{-i\sqrt{\pi}\Xi_\sigma({\bf r}_1)}\right\rangle_T+e^{\pi\mathcal{C}^\Xi_\sigma({\bf r}_1,{\bf r}')}
	\right],\label{eq:fvarphic}\\&&\langle f^\theta_c(x')\rangle_t=0,\label{eq:fthetac}\\
        &&\langle f^\varphi_s(x')\rangle_t=\frac{t^2_c\sqrt{2\pi}}{4}\delta(x'-x_0)
	\int d{\bf r}_1\delta(x_1-x_0)\nonumber\\
	&&\sum_{m,\sigma} \sigma me^{im\sqrt{\pi}\left( A^\Xi_\sigma({\bf r}')
	-A^{\Xi}_\sigma({\bf r}_1)\right)}\nonumber\\
	&&\left[ -\left\langle e^{i\sqrt{\pi}\Xi_\sigma({\bf r}')}
	e^{-i\sqrt{\pi}\Xi_\sigma({\bf r}_1)}\right\rangle_T+e^{\pi\mathcal{C}^\Xi_\sigma({\bf r}_1,{\bf r}')}
	\right]\nonumber\\
	&&+\frac{t^2_s\sqrt{2\pi}}{4}\delta(x'-x_0)
	\int d{\bf r}_1\delta(x_1-x_0)\nonumber\\
	&&\sum_{m,a} me^{im\sqrt{\pi}\left( A^\Psi_a({\bf r}')
	-A^{\Psi}_a({\bf r}_1)\right)}\left[ -\left\langle e^{i\sqrt{\pi}\Psi_a({\bf r}')}
	e^{-i\sqrt{\pi}\Psi_a({\bf r}_1)}\right\rangle_T\right.\nonumber\\
	&&\left.+e^{\pi\mathcal{C}^\Psi_a({\bf r}_1,{\bf r}')}
	\right],\label{eq:fvarphis}\\
	&&\langle f^\theta_s(x')\rangle_t=\frac{t^2_s\sqrt{2\pi}}{4}\delta(x'-x_0)
	\int d{\bf r}_1\delta(x_1-x_0)\nonumber\\
	&&\sum_{m,a} ma e^{im\sqrt{\pi}\left( A^\Psi_a({\bf r}')
	-A^{\Phi}_a({\bf r}_1)\right)}\nonumber\\
	&&\left[ -\left\langle e^{i\sqrt{\pi}\Psi_a({\bf r}')}
	e^{-i\sqrt{\pi}\Psi_a({\bf r}_1)}\right\rangle_T+e^{\pi\mathcal{C}^\Psi_a({\bf r}_1,{\bf r}')}
	\right],	\label{eq:fthetas}
\end{eqnarray}
where $\langle\ldots\rangle_T$~denotes the time ordered expectation value. In the latter equations we have used $e^Ae^B=e^{A+B}e^{\frac{1}{2}[A,B]}$, $\langle e^f\rangle_0=e^{\frac{1}{2}\langle
f^2\rangle_0}$, and  $\langle e^{if({\bf r}_1)}e^{-if({\bf r}_2)}\rangle_0
=\langle e^{-if({\bf r}_1)}e^{if({\bf r}_2)}\rangle_0=e^{\langle f({\bf r}_1)f({\bf r}_2)-f^2(0)\rangle_0}$.
$\bar\Xi_\sigma=\sqrt{2}(\bar\varphi_c+\sigma\bar\varphi_s)$, $\bar\Psi_a=\sqrt{2}(\bar\varphi_s+a\bar\theta_s)$,
$\mathcal{C}^\Xi_\sigma=\langle \Xi_\sigma\Xi_\sigma\rangle_0=2\left( \mathcal{C}_c+ \mathcal{C}_s\right)$
and $\mathcal{C}^\Psi_a=\langle \Phi_a\Phi_a\rangle_0=2\left[ \mathcal{C}_s+ \mathcal{D}_s+a\left(
\mathcal{F}_s+\mathcal{Q}_s\right)\right]$,
$A_{\Xi,\sigma}=\sqrt{2}\left( A^{\varphi}_c+\sigma A^{\varphi}_s \right)$
and $A_{\Psi,a}=\sqrt{2}\left(  A_s^\varphi+aA_s^\theta \right)$.

As an example, we show the calculation for $\langle f^\varphi_c(x')\rangle_t$ in some detail,
\begin{eqnarray}
	&&\langle f^\varphi_c(x')\rangle_t\approx\sum_\sigma it^2_c\sqrt{2\pi}\delta(x'-x_0)
	\int d{\bf r}_1\delta(x_1-x_0)\nonumber\\
	&&\sum_\eta \eta \left\langle \cos\left( \sqrt{2\pi}\bar\Xi^+_\sigma({\bf r}') \right)
	\sin\left( \sqrt{2\pi}\bar\Xi^\eta_\sigma({\bf r}_1) \right)\right\rangle_0\\
	&&=\sum_\sigma \frac{t^2_c\sqrt{2\pi}}{4}\delta(x'-x_0)
	\int d{\bf r}_1\delta(x_1-x_0)\nonumber\\
	&&\sum_\eta \eta \left\langle -e^{i\sqrt{\pi}\bar\Xi^+_\sigma({\bf r}')}
	e^{-i\sqrt{\pi}\bar\Xi^\eta_\sigma({\bf r}_1)}+e^{-i\sqrt{\pi}\bar\Xi^+_\sigma({\bf r}')}
	e^{i\sqrt{\pi}\bar\Xi^\eta_\sigma({\bf r}_1)}\right\rangle_0\nonumber\\
	&&=\sum_\sigma \frac{t^2_c\sqrt{2\pi}}{4}\delta(x'-x_0)
	\int d{\bf r}_1\delta(x_1-x_0)\nonumber\\
	&&\sum_m me^{im\sqrt{\pi}\left( A^\Xi_\sigma({\bf r}')
	-A^{\Xi}_\sigma({\bf r}_1)\right)}\nonumber\\
	&&\left[ -\left\langle e^{i\sqrt{\pi}\Xi_\sigma({\bf r}')}
	e^{-i\sqrt{\pi}\Xi_\sigma({\bf r}_1)}\right\rangle_T+e^{\pi\mathcal{C}_0({\bf r}_1,{\bf r}')}
	\right] . \nonumber
	\label{eq:fvarphi_emp}
\end{eqnarray}
Then with the help of the expressions for the correlation functions listed in the next section,
the second order correction to the spin density and charge current can be calculated as follows.
\begin{eqnarray}
	&&\langle \rho_{2,s}(x)\rangle={\rm sgn}(x-x_0)\int dt_1\nonumber\\
	&&\left[ \frac{t_c^2}{4v_f}
	\sum_{m,\sigma}m\sigma e^{im\sqrt{\pi}(A^\Xi_\sigma(x_0,t_1)-A^\Xi_\sigma(x_0,0))}
	e^{\pi\mathcal{C}^\Xi_\sigma(x_0,x_0,t_1)}\right.\nonumber\\
	&&\left.+\frac{t_s^2}{4v_f}\sum_{m,a}m
	e^{im\sqrt{\pi}\left( A^\Psi_a(x_0,t_1)-A^\Psi_a(x_0,0) \right)}e^{\pi\mathcal{C}^\Psi_a
	(x_0,x_0,t_1)}\right]+\nonumber\\
	&&\frac{t^2_s}{4v_f}\int dt_1\sum_{m,a}mae^{im\sqrt{\pi}
	(A^\Psi_a(x_0,t_1)-A^\Psi_a(x_0,0))}\nonumber\\
	&&e^{\pi\mathcal{C}^\Psi_a(x_0,x_0,t_1)}
    \label{eq:rhos2}
\end{eqnarray}
and
\begin{eqnarray}
    &&\langle j_{2,c}\rangle=-\frac{et_c^2}{4}\int dt_1
    \sum_{\sigma,m}m e^{im\sqrt{\pi}(A^\Xi_\sigma(x_0,t_1)-A^\Xi_\sigma(x_0,0))}\nonumber\\
    &&   e^{\pi\mathcal{C}^\Xi_\sigma(x_0,x_0,t_1)}
    \label{eq:jc2}
\end{eqnarray}
From the expressions for $A^{\Xi(\Psi)}_{\sigma(a)}$ in the next section, we find that
\begin{eqnarray}
	&&\langle \rho_{2,s}(x)\rangle=\frac{t^2_s}{4v_f}\int dt_1\nonumber\\
	&&\sum_{m,a}ma
	e^{imat_1eV_s}e^{\pi\mathcal{C}^\Psi_a(x_0,x_0,t_1)}
    \label{eq:rhos21}
\end{eqnarray}
and
\begin{eqnarray}
    &&\langle j_{2,c}\rangle=-\frac{et_c^2}{4}\int dt_1\nonumber\\
    &&   \sum_{\sigma,m}m e^{imt_1eV_c}
    e^{\pi\mathcal{C}^\Xi_\sigma(x_0,x_0,t_1)}
    \label{eq:jc21}
\end{eqnarray}
Numerical results obtained by the evaluation of the above integrals are presented in the main text.
For the limit $L\rightarrow\infty$, the correlation functions yield
\begin{eqnarray}
    &&\mathcal{C}^\Xi_\sigma(x_0,x_0,t)=-\frac{1}{2\pi}\left( g_0+\frac{1}{g_0} \right)\nonumber\\
    &&    \ln\left( \frac{(\tau_{\rm cu}+it\omega_L)^2}{\tau_{\rm cu}^2}\right)\\
    &&\mathcal{C}^\Psi_a(x_0,x_0,t)=-\frac{1}{2\pi}\left( g_0+\frac{1}{g_0} \right)\nonumber\\
    &&    \ln\left( \frac{(\tau_{\rm cu}+it\omega_L)^2}{\tau_{\rm cu}^2}\right),
    \label{eq:corfun1}
\end{eqnarray}
where $\omega_L=\frac{v_f}{g_0L}$ and $\tau_{cu}$ is the short time cut-off. 
In this case, the correction to the spin density and the charge current can be calculated analytically to give
\begin{eqnarray}
	\rho_{2,s}=\frac{\pi t^2_s\tau_{\rm cu}^{g_0+\frac{1}{g_0}}}{\hbar^2v_f\omega_L\Gamma\left( g_0+\frac{1}{g_0} \right)}
	{\rm sgn}\left( eV_s \right)\left|\frac{eV_s}{\hbar\omega_L}\right|^{g_0+\frac{1}{g_0}-1}
    \label{eq:Spinb1}
\end{eqnarray}
and
\begin{eqnarray}
	j_{2,c}=-\frac{e\pi t^2_c\tau_{\rm cu}^{g_0+\frac{1}{g_0}}}{\hbar^2\omega_L\Gamma\left( g_0+\frac{1}{g_0} \right)}
	{\rm sgn}\left( eV_c \right)\left|\frac{eV_c}{\hbar\omega_L}\right|^{g_0+\frac{1}{g_0}-1}, \nonumber\\
    \label{eq:currentb1}
\end{eqnarray}
where $\hbar$ is recovered. 

Next we consider a long junction, which is modeled as a spinful Tomonaga Luttinger liquid.
As shown in the main text, in the region $x<|d/2|$ where $d$ is the length of the junction,
the Hamiltonian is given by $\hat{H}_t=\hat{H}^c_{t0}+\hat{H}^s_{t0}
+\hat{H}_{t2}$. We find that $\hat{H}^c_{t0}=\frac{\hbar v_f}{2}\int dx\left[ (\partial_x\theta_c)^2+\frac{1}{g_0^2}
    (\partial_x\varphi_c)^2\right]$, so that the Hamiltonian of the charge sector has a constant interaction parameter $g_0$~ from $-L/2$~ to $+L/2$. Therefore there is no scattering in the charge sector. However, for the spin sector, not only the interaction parameter is changed across the junction
    ($\hat{H}^s_{t0}=\frac{\hbar v_f}{2}\int dx\left[ (\partial_x\theta_s)^2+\frac{1}{g_{s}^2}(\partial_x\varphi_s)^2\right]$),
    but there is also an additional term $\hat{H}_{t2}$ \cite{gritsev2005,tanaka2009}
\begin{eqnarray}
    \hat{H}_{t2}=\int dx\left[ g_{1\perp}\cos\left( \sqrt{8\pi}\varphi_s \right)
    +g_{sf}\cos\left( \sqrt{8\pi}\theta_s \right)\right],\nonumber\\
    \label{eq:Ht2}
\end{eqnarray}
which comes from two particle scattering.
In order to focus on the new physics brought about by the long junction during our perturbative calculation, we assume a hierarchy of length scales $L \gg d \gg \lambda_f~$ which simplifies the problem.
We can then take the $L\rightarrow \infty$ limit, which amounts to neglecting the influence of the leads. This is a reasonable assumption here because it is shown in the main text that already the finite size corrections with respect to
$d$~ are small compared to the leading $d \rightarrow \infty~$ contribution. Hence, finite size corrections with respect to $L$~are negligible.
Furthermore, we assume that no impurities exist in the junction region, so that single particle
backscattering can be neglected. This corresponds to the physical situation of a clean quantum wire.
Within the above approximations, our model again becomes a $g(x)$ model so that the formalism (\ref{eq:rho_1})
-(\ref{eq:current_3}) can be applied.

With the Hamiltonian (\ref{eq:Ht2}), we have
\begin{eqnarray}
    \hat{f}^{\varphi}_{s}=g_{1\perp}\sqrt{8\pi}\sin\sqrt{8\pi}\left( \varphi^+_s
    +A^{+,\varphi}_s\right)\\
    \hat{f}^{\theta}_{s}=g_{sf}\sqrt{8\pi}\sin\sqrt{8\pi}\left( \theta^+_s
    +A^{+,\theta}_s\right)
    \label{eq:fs2}
\end{eqnarray}

Then up to the first order term we get
\begin{eqnarray}
	&&\langle \hat{f}^{\varphi}_{s}({\bf r}')\rangle_t=\sqrt{\frac{\pi}{2}}g_{1\perp}^2\int d{\bf r}_1
	\sum_mme^{i\sqrt{8\pi}m\left( A^\varphi_s({\bf r}')-A^\varphi_s({\bf r}_1) \right)}\nonumber\\
	&&\left[
	-\langle e^{i\sqrt{8\pi}\varphi_s({\bf r}')}e^{-i\sqrt{8\pi}\varphi_s({\bf r}_1)}\rangle_T+
	e^{8\pi\mathcal{C}_s({\bf r}_1,{\bf r}')}\right]\\
    	&&\langle \hat{f}^{\theta}_{s}({\bf r}')\rangle_t=\sqrt{\frac{\pi}{2}}g_{sf}^2\int d{\bf r}_1
	\sum_mme^{i\sqrt{8\pi}m\left( A^\theta_s({\bf r}')-A^\theta_s({\bf r}_1) \right)}\nonumber\\
	&&\left[
	-\langle e^{i\sqrt{8\pi}\theta_s({\bf r}')}e^{-i\sqrt{8\pi}\theta_s({\bf r}_1)}\rangle_T+
	e^{8\pi\mathcal{D}_s({\bf r}_1,{\bf r}')}\right]
    \label{eq:fsavg2}
\end{eqnarray}
It is convenient to divide $\rho_{2,s}$ into two parts
\begin{eqnarray}
    &&\langle \rho_{2,s}(\bold{r})\rangle=\langle \rho^{\varphi}_{2,s}(\bold{r})\rangle
    +\langle \rho^{\theta}_{2,s}(\bold{r})\rangle
        \label{eq:rhosde1}
\end{eqnarray}
with
\begin{eqnarray}
	&&\langle \rho^{\varphi}_{2,s}(\bold{r})\rangle=\frac{g^2_{1\perp}\tilde{g}_s}{2\omega^s_dd}\int dx'dx_1dt_1
    {\rm sgn}(x-x')\nonumber\\
    &&\sum_{m}me^{im\sqrt{8\pi}(A^{\varphi}_s(x',0)-A^{\varphi}_s(x_1,t_1))}\nonumber\\
    &&    \left[ \langle e^{i\sqrt{8\pi}\varphi_s(x',0)}e^{-i\sqrt{8\pi}\varphi_s(x_1,r_1)}\rangle_T
    -e^{8\pi\mathcal{C}_s(x_1,x',t_1)}\right]\nonumber\\
    &&=\frac{g^2_{1\perp}\tilde{g}_s}{2\omega^s_dd}\int dx'dx_1dt_1{\rm sgn}(x-x')\nonumber\\
    &&\sum_{m}me^{-i2m\frac{g_{s}^2g_0}{v_f}(x_1-x')eV_s}\nonumber\\
    &&   \left[ \langle e^{i\sqrt{8\pi}\varphi_s(x',0)}e^{-i\sqrt{8\pi}\varphi_s(x_1,r_1)}\rangle_T\right.\nonumber\\
     &&\left.-e^{8\pi\mathcal{C}_s(x_1,x',t_1)}\right]\label{eq:rhosvarphi}\\
    &&\langle \rho^{\theta}_{2,s}(\bold{r})\rangle=\frac{g^2_{sf}\tilde{g}_s}{2\omega_d^sd}\int dx'dx_1dt_1\nonumber\\
    &&   \sum_{m}me^{im\sqrt{8\pi}(A^{\theta}_s(x',0)-A^{\theta}_s(x_1,t_1))}\nonumber\\
    &&    \left[ \langle e^{i\sqrt{8\pi}\theta_s(x',0)}e^{-i\sqrt{8\pi}\theta_s(x_1,t_1)}\rangle_T
    -e^{8\pi\mathcal{D}_s(x_1,x',t_1)}\right]\nonumber\\
    &&=-\frac{g^2_{sf}\tilde{g}_s}{2\omega_d^sd}\int dx'dx_1dt_1\nonumber\\
    &&    \sum_{m}me^{-i2mt_1eV_s}   e^{8\pi\mathcal{D}_s(x_1,x',t_1)}
        \label{eq:rhostheta}
\end{eqnarray}
Here we have taken $\mu_{1,s}+\mu_{2,s}=0$ and $\tilde{g}_s=g_sg_0$. If we consider the region $x>|d/2|$, then it is easy to show
that (\ref{eq:rhosvarphi}) vanishes because $\sum_mme^{i2m\frac{g_{s}^2g_0}{v_f}(x'-x_1)eV_s}$ is odd in $r=x'-x_1$ while
$e^{8\pi\mathcal{C}_{0s}(x_1,0,x',t')}$ is even in $r$. Eq. (\ref{eq:rhostheta}) can now be
rewritten as
\begin{eqnarray}
	&&\langle \rho^{\theta}_{2,s}\rangle=-\frac{g_{sf}^2\tilde{g}_s}{2(\hbar\omega_d^s)^2d}\int^{d/2}_{-d/2}dx'
	\int^{d/2}_{-d/2}dx_1 f(x',x_1)\nonumber\\
	&&=-\frac{(g_{sf}d)^2\tilde{g}_s}{4(\hbar\omega_d^s)^2d}\int^{1}_{-1}dR
	\int^{1-|R|}_{-1+|R|}dr f(r,R)\\
	&&f(r,R)=\int^{\infty}_{-\infty}d\tau_1\sum_m m e^{-i2m\tau_1\frac{eV_s}{\hbar\omega^s_d}}\nonumber\\
	&&e^{8\pi\mathcal{D}_{0s}\left( r,R,\tau_1 \right)}
	\label{eq:rhostheta1}
\end{eqnarray}
where $r=\frac{x_1-x'}{d}$, $R=\frac{x_1+x'}{d}$
and $\tau_1=t_1\omega^s_d$ with $\omega^s_d=\frac{v_f}{g_{s}d}$. The full correlation function
$\mathcal{D}_{s}\left( r,R,\tau_1 \right)$ can be found in the next section.
For comparison with the above single particle scattering process,
we also divide $\rho^{\theta}_{2,s}$ into two parts
$\rho^{\theta}_{2,s}=\rho^{\theta}_{2,s1}+\rho^{\theta}_{2,s2}$.
The first part $\rho^{\theta}_{2,s1}$ is the contribution for a quite long junction
so that we can use the correlation function of the infinite wire, given by
$\mathcal{D}_{s}\left( r,R,\tau \right)=-\frac{1}{4\pi\tilde{g}_s}\ln
	\left( \frac{(\tau_{\rm cu}+i\tau)^2+r^2)}{\tau_{\rm cu}^2} \right)$ which is
independent of $R$. Then we can easily integrate over R. Furthermore
all the singularities occurring during the integrations appear in
$\rho^{\theta}_{2,s1}$, so that our separate analysis of $\rho^{\theta}_{2,s1}$ can help
to keep the computing time of the whole calculation sustainable. In the infinite wire part $\rho^{\theta}_{2,s1}$,
$f(r)$ is independent of $R$. Thus the expression (\ref{eq:rhostheta1})
can be simplified to
\begin{eqnarray}
	&&\rho^{\theta}_{2,s1}=-\frac{(g_{sf}d)^2\tilde{g}_s}{2(\hbar\omega^s_d)^2d}\int^{1}_{-1}dr
	(1-|r|) f(r)\nonumber\\
	&&=\frac{(g_{sf}d)^2\tilde{g}_s}{2(\hbar\omega^s_d)^2d}\int^{1}_{-1}dr\int^{\infty}_0 d\tau'(-4)\left( 1-|r| \right)
	\sin(2\frac{eV_s}{\hbar\omega^s_d}\tau')\nonumber\\
	&&e^{8\pi\Re\mathcal{D}_{s}(r,\tau')}\sin8\pi\Im\mathcal{D}_{s}(r,\tau').
	\label{eq:hlsll_rhotheta2}
\end{eqnarray}
This can be further simplified with the help of the following equality
\begin{eqnarray}
	&&\int^{\infty}_{-\infty}dx \left( a+i(x+b) \right)^{-\mu} \left( a+i(x-b) \right)^{-\mu}
	e^{ipx}=\nonumber\\
	&&\theta(p)\frac{2\pi^{3/2}e^{-ap}}{\Gamma(\mu)}\left( \frac{2|b|}{p} \right)^{\frac{1}{2}-\mu}
	J_{\mu-\frac{1}{2}}(|b|p)
	\label{eq:intequality}
\end{eqnarray}
Here we require $a>0$ and $\mu>1/2$. With the above equality, Eq.~(\ref{eq:hlsll_rhotheta2})
can be calculated to yield
\begin{eqnarray}
	&&\rho^{\theta}_{2,s1}=\frac{\pi^{3/2}\tau_{cu}^{4/\tilde{g}_s}(g_{sf}d)^2\tilde{g}_s}
	{\Gamma(2/\tilde{g}_s)(\hbar\omega^s_d)^2d}sgn\left( eV_s \right)\left|\frac{eV_s}{\hbar\omega^s_d}\right|^{\frac{2}{\tilde{g}_s}-\frac{1}{2}}
	\nonumber\\
	&&\int^1_{-1}dr(1-|r|)|r|^{\frac{1}{2}-\frac{2}{\tilde{g}_s}}J_{\frac{2}{\tilde{g}_s}-\frac{1}{2}}(|2\frac{eV_s}{\hbar\omega^s_d}r|)
	\label{eq:hlsll_rhotheta3}
\end{eqnarray}
Similar to the calculation in Ref.~[\onlinecite{peguiron2007}], in the limit $eV_s/\hbar\omega_d\gg 1$,
the above expressions can be approximated as
\begin{eqnarray}
	&&\rho^{\theta}_{2,s1}\approx sgn\left( eV_s \right)
	\frac{\pi^2\tau_{cu}^{4/\tilde{g}_s}(g_{sf}d)^2\tilde{g}_s}{\Gamma^2(2/\tilde{g}_s)(\hbar\omega^s_d)^2d}
	\left|\frac{eV_s}{\hbar\omega^s_d}\right|^{\frac{4}{\tilde{g}_s}-2} .\nonumber\\
	\label{eq:hlsll_rhotheta4}
\end{eqnarray}

\section{Correlation functions}
In this section, we will provide some useful relations for the correlation functions used in the main text.
First, we give a list of the zero temperature expressions for the correlation functions $\mathcal{C}$, $\mathcal{F}$, $\mathcal{Q}$, and $\mathcal{D}$ (derived in a similar way as in Ref.~[\onlinecite{dolcini2005}]).
\begin{widetext}
\begin{eqnarray}
	&&\mathcal{C}(\xi,\eta,\tau)=-\frac{\tilde{g}}{4\pi}\left\{\sum_{m\in Z_{even}}\tilde\gamma^{|m|}\ln\left(
	\frac{(\tau_{\rm cu}+i\tau)^2+(\xi_r+m)^2}{\tau_{\rm cu}^2+m^2}\right)+\sum_{m\in Z_{odd}}\tilde\gamma^{|m|}\ln\left(
	\frac{(\tau_{\rm cu}+i\tau)^2+(\xi_R+m)^2}{\tau_{\rm cu}^2+(\xi_R+m)^2}\right)\right.\nonumber\\
	&&\left.+\frac{1}{2}\sum_{m\in Z_{odd}}\tilde\gamma^{|m|}
	\ln\left( \frac{[\tau_{\rm cu}^2+(\xi_R+m)^2]^2}{[\tau_{\rm cu}^2+(2\xi+m)^2][\tau_{\rm cu}^2+(2\eta+m)^2]} \right)\right\},\\
	&&\mathcal{D}(\xi,\eta,\tau)=-\frac{1}{4\pi \tilde{g}}\left\{\sum_{m\in Z_{even}}\tilde\gamma^{|m|}\ln\left(
	\frac{(\tau_{\rm cu}+i\tau)^2+(\xi_r+m)^2}{\tau_{\rm cu}^2+m^2}\right)-\sum_{m\in Z_{odd}}\tilde\gamma^{|m|}\ln\left(
	\frac{(\tau_{\rm cu}+i\tau)^2+(\xi_R+m)^2}{\tau_{\rm cu}^2+(\xi_R+m)^2}\right)\right.\nonumber\\
	&&\left.-\frac{1}{2}\sum_{m\in Z_{odd}}\tilde\gamma^{|m|}
	\ln\left( \frac{[\tau_{\rm cu}^2+(\xi_R+m)^2]^2}{[\tau_{\rm cu}^2+(2\xi+m)^2][\tau_{\rm cu}^2+(2\eta+m)^2]} \right)\right\},\\
	&&\mathcal{Q}(\xi,\eta,\tau)=-\frac{1}{4\pi}\left\{-\sum_{m\in Z_{even}}\tilde\gamma^{|m|}\ln\left(
	\frac{(\tau_{\rm cu}+i\tau)+i(\xi_r+m)}{(\tau_{\rm cu}+i\tau)-i(\xi_r+m)}\right)-\sum_{m\in Z_{odd}}\tilde\gamma^{|m|}\ln\left(
	\frac{(\tau_{\rm cu}+i\tau)+i(\xi_R+m)}{(\tau_{\rm cu}+i\tau)-i(\xi_R+m)}\right)\right\}, \nonumber\\\\
	&&\mathcal{F}(\xi,\eta,\tau)=-\frac{1}{4\pi}\left\{-\sum_{m\in Z_{even}}\tilde\gamma^{|m|}\ln\left(
	\frac{(\tau_{\rm cu}+i\tau)+i(\xi_r+m)}{(\tau_{\rm cu}+i\tau)-i(\xi_r+m)}\right)+\sum_{m\in Z_{odd}}\tilde\gamma^{|m|}\ln\left(
	\frac{(\tau_{\rm cu}+i\tau)+i(\xi_R+m)}{(\tau_{\rm cu}+i\tau)-i(\xi_R+m)}\right)\right\}. \nonumber\\
	\label{eq:correlation}
\end{eqnarray}
\end{widetext}
In the above expressions, the parameters depend on the details of the configuration under consideration.
For the short junction case, $\xi=x/L$, $\eta=y/L$, $\xi_r=\xi-\eta$, $\xi_R=\xi+\eta$,
$\tau=t\omega_L$, $\omega_L=\frac{v_f}{g_0L}$ and $\tau_{\rm cu}$ is the short time cut-off.
$\tilde{g}$ and $\tilde{\gamma}=\frac{1-\tilde{g}}{1+\tilde{g}}$ are different for charge and spin sector. $\tilde{g}=\tilde{g}_c=g_0$ for the charge sector while $\tilde{g}=\tilde{g}_s=1/g_0$
for the spin sector.
For the long junction, $\xi=x/d$, $\eta=y/d$, $\xi_r=\xi-\eta$, $\xi_R=\xi+\eta$,
$\tau=t\omega_d$ and $\tau_{\rm cu}$ is the short time cut-off, where $d$ is the length
of the junction. Now, $\omega_d=\omega^c_d=\frac{v_f}{g_0d}$ and $\tilde{g}=\tilde{g}_c=1$ for the charge sector,
while $\omega_d=\omega^s_d=\frac{v_f}{g_{s}d}$ and $\tilde{g}=\tilde{g}_s=g_{s}g_0$ for the spin sector.

Besides the above non-time ordered correlation functions, we have three other types of correlation functions:
the retarded correlation function, the advanced correlation function, and the Keldysh correlation function
defined in App. A. Useful identities for these correlation functions are
\begin{eqnarray}
    &&\int^{\infty}_{-\infty}d\tau'e^{i\bar\omega (\tau-\tau')}\partial_\tau\mathcal{C}^R(\xi,\eta,\tau-\tau')
    =\nonumber\\
    &&-i\frac{\tilde{g}}{2}\left( e^{i\bar\omega |\xi_r|}+\frac{2(\tilde\gamma^2e^{i2\bar\omega}\cos(\bar\omega \xi_r)
    +\tilde\gamma e^{i\bar\omega}\cos(\bar\omega \xi_R))}{1-\tilde\gamma^2e^{i2\bar\omega}} \right)\nonumber\\\\
    &&\int^{\infty}_{-\infty}d\tau'e^{i\bar\omega (\tau-\tau')}\partial_\xi\mathcal{C}^R(\xi,\eta,\tau-\tau')
    =\nonumber\\
    &&i\frac{\tilde{g}}{2}{\rm sgn}(\xi_r)e^{i\bar\omega |\xi_r|}-\frac{\tilde{g}(\tilde\gamma^2e^{i2\bar\omega}\sin(\bar\omega \xi_r)
    +\tilde\gamma e^{i\bar\omega}\sin(\bar\omega\xi_R))}{1-\tilde\gamma^2e^{i2\bar\omega}}\nonumber\\
   \label{eq:cf_Ctx}
\end{eqnarray}
\begin{eqnarray}
 &&\int^{\infty}_{-\infty}d\tau'e^{i\bar\omega (\tau-\tau')}\partial_\tau\mathcal{Q}^R(\xi,\eta,\tau-\tau')
    \nonumber\\
    &&=\frac{i}{2}{\rm sgn}(\xi_r)\left( 1-e^{i\bar\omega |\xi_r|} \right) + \frac{\tilde\gamma^2e^{i2\bar\omega}\sin(\bar\omega\xi_r)
    +\tilde\gamma e^{i\bar\omega}\sin(\bar\omega\xi_R)}{1-\tilde\gamma^2e^{i2\bar\omega}}\nonumber\\\\
    &&\int^{\infty}_{-\infty}d\tau'e^{i\bar\omega (\tau-\tau')}\partial_\xi\mathcal{Q}^R(\xi,\eta,\tau-\tau')\nonumber\\
    &&=\frac{i}{2}e^{i\bar\omega |\xi_r|} + i \frac{\tilde\gamma^2e^{i2\bar\omega}\cos(\bar\omega \xi_r)
    +\gamma e^{i\bar\omega}\cos(\bar\omega \xi_R)}{1-\tilde\gamma^2e^{i2\bar\omega}}
    \label{eq:cf_Qtx}
\end{eqnarray}
\begin{eqnarray}
    &&\int^{\infty}_{-\infty}d\tau'e^{i\bar\omega (\tau-\tau')}\partial_\tau\mathcal{F}^R(\xi,\eta,\tau-\tau')\nonumber\\
    &&=\frac{i}{2}{\rm sgn}(\xi_r)\left( 1-e^{i\bar\omega |\xi_r|} \right)+\frac{(\tilde\gamma^2e^{i2\bar\omega}\sin(\bar\omega\xi_r)
    -\tilde\gamma e^{i\bar\omega}\sin(\bar\omega\xi_R))}{1-\tilde\gamma^2e^{i2\bar\omega}}\nonumber\\\\
    &&\int^{\infty}_{-\infty}d\tau'e^{i\bar\omega (\tau-\tau')}\partial_\xi\mathcal{F}^R(\xi,\eta,\tau-\tau')\nonumber\\
    &&=\frac{i}{2}e^{i\bar\omega |\xi_r|}+i\frac{(\tilde\gamma^2e^{i2\bar\omega}\cos(\bar\omega \xi_r)
    -\tilde\gamma e^{i\bar\omega}\cos(\bar\omega\xi_R))}{1-\tilde\gamma^2e^{i2\bar\omega}}
    \label{eq:cf_Ftx}
\end{eqnarray}
\begin{eqnarray}
    &&\int^{\infty}_{-\infty}d\tau'e^{i\bar\omega (\tau-\tau')}\partial_\tau\mathcal{D}^R(\xi,\eta,\tau-\tau')\nonumber\\
    &&=-\frac{i}{2\tilde{g}}\left( e^{i\bar\omega |\xi_r|}+\frac{2(\tilde\gamma^2e^{i2\bar\omega}\cos(\bar\omega\xi_r)
    -\tilde\gamma e^{i\bar\omega}\cos(\bar\omega\xi_R))}{1-\tilde\gamma^2e^{i2\bar\omega}} \right)\nonumber\\\\
    &&\int^{\infty}_{-\infty}d\tau'e^{i\bar\omega (\tau-\tau')}\partial_\xi\mathcal{D}^R(\xi,\eta,\tau-\tau')\nonumber\\
    &&=\frac{i}{2\tilde{g}}{\rm sgn}(\xi_r)e^{i\bar\omega |\xi_r|}-\frac{\tilde\gamma^2e^{i2\bar\omega}\sin(\bar\omega\xi_r)
    -\tilde\gamma e^{i\bar\omega}\sin(\tilde\omega \xi_R)}{\tilde{g}(1-\tilde\gamma^2e^{i2\bar\omega})}\nonumber\\
    \label{eq:cf_Dtx}
\end{eqnarray}

As seen in Sec.~II, besides the correlation functions we also need
$A^{\varphi(\theta)}_\lambda(\bold{r}_1)-A^{\varphi(\theta)}_\lambda(\bold{r}')$,
where $\lambda=c,s$ for our calculations, which are given by
\begin{eqnarray}
    &&A^\varphi_\lambda(\bold{r}_1)-A^\varphi_\lambda(\bold{r}')\nonumber\\
    &&=-\frac{i}{\sqrt{2\pi}}\int d\bold{r}_2\left[\left( \int^{t_1}_{t'} dt_3\partial_{t_3}\mathcal{C}^R_\lambda
    (x_1,t_3,\bold{r}_2)\right.\right.\nonumber\\
    &&\left.\left.+\int^{x_1}_{x'}dx_3 \partial_{x_3}\mathcal{C}^R_\lambda(x_3,t',\bold{r}_2)\right)
    \partial_{x_2}\mu^\varphi_\lambda(\bold{r}_2)\right.\nonumber\\
    &&\left.+\left( \int^{t_1}_{t'} dt_3\partial_{t_3}\mathcal{F}^R_\lambda
    (x_1,t_3,\bold{r}_2) +\int^{x_1}_{x'}dx_3 \partial_{x_3}\mathcal{F}^R_\lambda(x_3,t',\bold{r}_2)\right)
   \right.\nonumber\\
   &&\left.\partial_{x_2}\mu^\theta_\lambda(\bold{r}_2)\right]
    \label{eq:cf_A}
\end{eqnarray}
as well as
\begin{eqnarray}
    &&A^\theta_\lambda(\bold{r}_1)-A^\theta_\lambda(\bold{r}')\nonumber\\
    &&=-\frac{i}{\sqrt{2\pi}}\int d\bold{r}_2\left[\left( \int^{t_1}_{t'} dt_3
    \partial_{t_3}\mathcal{Q}^R_\lambda(x_1,x_2,t_3-t_2)\right.\right.\nonumber\\
    &&\left.+\int^{x_1}_{x'}dx_3
    \partial_{x_3}\mathcal{Q}^R_\lambda(x_3,x_2,t'-t_2)\right)
    \partial_{x_2}\mu^\varphi_\lambda(x_2)\nonumber\\
    &&+\left( \int^{t_1}_{t'} dt_3
    \partial_{t_3}\mathcal{D}^R_\lambda(x_1,x_2,t_3-t_2)\right.\nonumber\\
    &&\left.\left.+\int^{x_1}_{x'}dx_3
    \partial_{x_3}\mathcal{D}^R_\lambda(x_3,x_2,t'-t_2)\right)
    \partial_{x_2}\mu^\theta_\lambda(x_2)\right].
    \label{eq:cf_A1}
\end{eqnarray}
Since for the charge part only $\mu_c^\varphi$ contributes, we have
\begin{eqnarray}
    &&A^\varphi_c(\bold{r}_1)-A^\varphi_c(\bold{r}')\nonumber\\
    &&=\frac{1}{2\sqrt{2\pi}}\int dx_2\left( -(t_1-t')\right.\nonumber\\
    &&\left.+\frac{\tilde{g}}{\omega^c_LL}\left( |x_1-x_2|-|x'-x_2| \right)\right)
    \partial_{x_2}\mu_c(x_2)\nonumber\\
    &&=\frac{1}{2\sqrt{2\pi}}\left[ (t_1-t')(\mu_{1,c}-\mu_{2,c})\right.\nonumber\\
    &&\left.-\frac{\tilde{g}_c}{\omega^c_LL}(\mu_{1,c}+\mu_{2,c})(x_1-x') \right]\\
    &&A^\theta_c(\bold{r}_1)-A^\theta_c(\bold{r}')\nonumber\\
    &&=-\frac{i}{\sqrt{2\pi}}\int dx_2\left( \int^{x_1}_{x'}dx_3 \frac{i}{2\omega^c_LL\tilde{g}_c} \right)
    \partial_{x_2}\mu_c(x_2)\nonumber\\
    &&=-\frac{(x_1-x')(\mu_{1,c}-\mu_{2,c})}{2\sqrt{2\pi}\tilde{g}_c\omega^c_LL},
    \label{eq:cf_A2}
\end{eqnarray}
whereas for spin part, only $\mu_s^\theta$ contributes, yielding
\begin{eqnarray}
    &&A^\theta_s(\bold{r}_1)-A^\theta_s(\bold{r}')
    =-\frac{i}{\sqrt{2\pi}}\int d\bold{r}_2\nonumber\\
    &&\left( \int^{t_1}_{t'} dt_3
    \partial_{t_3}\mathcal{D}^R_s(x_1,x_2,t_3-t_2)\right.\nonumber\\
    &&\left.+\int^{x_1}_{x'}dx_3
    \partial_{x_3}\mathcal{D}^R_s(x_3,x_2,t'-t_2)\right)
    \partial_{x_2}\mu_s(x_2)\nonumber\\
    &&=\frac{1}{2\sqrt{2\pi}}\left[ (t_1-t')(\mu_{1,s}-\mu_{2,s})\right.\nonumber\\
    &&\left.-\frac{1}{\omega^s_LL\tilde{g}_s}(\mu_{1,s}+\mu_{2,s})(x_1-x')\right]\\
    &&A^\varphi_s(\bold{r}_1)-A^\varphi_s(\bold{r}')
    =-\frac{i}{\sqrt{2\pi}}\int d\bold{r}_2\nonumber\\
    &&\left( \int^{t_1}_{t'} dt_3
    \partial_{t_3}\mathcal{F}^R_s(x_1,x_2,t_3-t_2)\right.\nonumber\\
    &&\left.+\int^{x_1}_{x'}dx_3
    \partial_{x_3}\mathcal{F}^R_s(x_3,x_2,t'-t_2)\right)
    \partial_{x_2}\mu_s(x_2)\nonumber\\
    &&=-\frac{\tilde{g}_s}{2\sqrt{2\pi}\omega^s_LL}\left( \mu_{1,s}-\mu_{2,s} \right)
    \left( x_1-x' \right).
    \label{eq:cf_A3}
\end{eqnarray}
For the short junction, using the relation between $\Xi$, $\Psi$ and $\varphi$, $\theta$, we obtain
\begin{eqnarray}
    A_{\Xi,\sigma}(x_0,t_1)-A_{\Xi,\sigma}(x_0,t')=\frac{1}{\sqrt{\pi}}(t_1-t')eV_c ,\\
    A_{\Psi,a}(x_0,t_1)-A_{\Psi,a}(x_0,t')=\frac{a}{\sqrt{\pi}}(t_1-t')eV_s .
\end{eqnarray}
For the long junction, we only need to change the parameters $L$ to $d$ and $\omega_L$ to $\omega_d$.

\bibliography{helical}

\end{document}